\documentclass[10pt]{article}
\usepackage{amsmath, amssymb,bbm}
\usepackage[all,cmtip]{xy}

\begin{document}
 \rightline{LTH 933} 
\vskip 1.5 true cm  
\begin{center}  
{\large The Hesse potential, the c-map and black hole solutions}\\[.5em]
\vskip 1.0 true cm   
{T.\ Mohaupt$^1$ and O.\ Vaughan$^{1,2}$} \\[3pt] 
$^1${Department of Mathematical Sciences\\ 
University of Liverpool\\
Peach Street \\
Liverpool L69 7ZL, UK\\  
Thomas.Mohaupt@liv.ac.uk, Owen.Vaughan@liv.ac.uk \\[1em]  
$^2${   Department of Mathematics\\  
and Center for Mathematical Physics\\ 
University of Hamburg\\ 
Bundesstra{\ss}e 55, 
D-20146 Hamburg, Germany\\  
owen.vaughan@math.uni-hamburg.de}
}\\[1em] 
December 13, 2011, revised January 19, 2012, revised June 13, 2012
\end{center}  
\vskip 1.0 true cm  
\baselineskip=18pt  
\begin{abstract}  
\noindent  
We present a new formulation of the local $c$-map, which makes
use of a symplectically covariant  real formulation of special 
K\"ahler geometry. We obtain an explicit and simple expression 
for the resulting quaternionic, or, in the case of reduction 
over time, para-quaternionic K\"ahler metric in terms of the
Hesse potential, which is similar
to the expressions for the metrics obtained from the 
rigid $r$- and $c$-map, and from the local $r$-map. 

As an application we use the
temporal version of the $c$-map to derive the black hole
attractor equations from geometric properties of the scalar
manifold, without imposing supersymmetry or spherical symmetry.
We observe that for general (non-symmetric) $c$-map spaces
static BPS solutions are related to a canonical family of
totally isotropic, totally geodesic submanifolds. Static non-BPS
solutions can be obtained by applying a field rotation matrix
which is subject to a non-trivial compatibility condition. We show that
for a class of prepotentials, which includes
the very special (`cubic') prepotentials as a subclass,  axion-free
solutions always admit a non-trivial field rotation matrix.

\end{abstract}


\newpage
 
\tableofcontents

\section{Introduction}

The special K\"ahler geometry of ${\cal N}=2$ vector multiplets
\cite{deWit:1984pk}
plays a central role in the study of the non-perturbative
properties of gauge theories \cite{Seiberg:1994rs,Seiberg:1994aj}, 
string compactifications \cite{Kachru:1995wm,Ferrara:1995yx,Kachru:1995fv}, 
and of
black holes, in particular the attractor mechanism \cite{Ferrara:1995ih},
black hole entropy \cite{Strominger:1996sh,Maldacena:1997de,LopesCardoso:1998wt,LopesCardoso:2000qm} and the OSV conjecture \cite{Ooguri:2004zv,LopesCardoso:2004xf,LopesCardoso:2006bg}. Its distinguished
feature is the existence of a single holomorphic function,
the prepotential, which encodes all vector multiplet couplings.
The power of holomorphicity is a key property, which sets
${\cal N}=2$ theories apart from ${\cal N}=1$ theories where
the K\"ahler potential is not related to an underlying 
holomorphic function. 
While at first glance our knowledge of special K\"ahler geometry
appears to be comprehensive, there are still aspects which deserve 
further study. 

\subsection{Projective special K\"ahler geometry in real
coordinates}

It is known that effective supergravity actions are subject to 
non-holomorphic corrections \cite{Dixon:1990pc}, which enter into 
the relation between the supergravity effective action and string 
amplitudes. This has 
consequences for black hole entropy and the 
OSV conjecture \cite{LopesCardoso:1999cv,LopesCardoso:1999ur,LopesCardoso:2004xf,LopesCardoso:2006bg,Cardoso:2008fr}. 
In this context it became clear that it is sometimes preferable to formulate
special K\"ahler geometry in terms of special real instead of
special holomorphic coordinates \cite{Freed:1997dp,LopesCardoso:2006bg}. 
This real formulation has been used to develop a manifestly 
duality covariant approach to the OSV conjecture  
\cite{LopesCardoso:2006bg,Cardoso:2006xz,Cardoso:2010gc}.

While the real formulation of the affine special K\"ahler geometry
of rigid vector multiplets is straightforward, the real formulation 
of the projective special K\"ahler geometry of local vector multiplets
leaves room for improvements. For affine special K\"ahler manifolds $N$
the special real coordinates are Darboux coordinates, and
the special K\"ahler metric is Hessian \cite{Freed:1997dp}. 
The Hesse potential is obtained by applying a Legendre transformation 
to the imaginary part of the prepotential \cite{Cortes:2001}. 
Electric-magnetic duality, which is a 
central feature of $N=2$ vector multiplets,
acts by symplectic transformations. While
the prepotential is not a symplectic function, the Hesse potential is,
and the special real coordinates form a symplectic vector. 
In \cite{Ferrara:2006at} 
a real formulation of projective special K\"ahler geometry
was worked out, and it was shown that only part of the symplectic
covariance of the underlying affine manifold could be kept manifest.
However, in applications such as black hole solutions and the 
study of non-holomorphic corrections one would like to have 
the full symplectic covariance manifest.

In this paper we obtain a real formulation 
of projective special K\"ahler geometry which is symplectically 
covariant. We make use of the superconformal formalism which 
employs the gauge equivalence between a theory of $n+1$ 
superconformal vector multiplets with scalar manifold $N$
and a theory of $n$ vector multiplets coupled to Poincar\'e
supergravity, with scalar manifold $\bar{N}=N/\mathbbm{C}^*=M//U(1)$, 
see for example \cite{Mohaupt:2011ab} for a review.
The main idea is to keep the $U(1)$ gauge invariance of the
superconformal formulation intact, which amounts to working
on $N$ or on the associated Sasakian $S$, which is a $U(1)$ principal
bundle over $\bar{N}$, instead of working on $\bar{N}$ itself. 
We derive explicit expressions for the scalar and vector
kinetic terms as real symmetric tensor fields on $N$.
These tensor fields can be expressed in terms of the Hesse potential and
are related to one another and to the metric
of the associated superconformal theory by adding differentials
dual to the vector fields generating the $\mathbbm{C}^*$-action.

\subsection{The $c$-map}

The special geometries of five-dimensional vector multiplets 
\cite{Gunaydin:1983bi},
four-dimensional vector multiplets \cite{deWit:1984pk} 
and of hypermultiplets \cite{Bagger:1983tt}
are related to one another by dimensional reduction. The
corresponding maps between the scalar manifolds are called
the $r$-map and the $c$-map respectively \cite{Cecotti:1988qn,Ferrara:1989ik,deWit:1991nm,deWit:1992wf}. Both maps have
rigid and local versions, depending on whether rigid or local
supersymmetry is considered. 
Moreover, by reducing over time
rather than space one obtains `temporal' versions
of the $r$- and $c$-map 
\cite{Cortes:2003zd,Cortes:2005uq,Gunaydin:2005mx,Neitzke:2007ke,Cortes:2009cs}, which
can be used for generating stationary solitonic solutions
by lifting Euclidean, instantonic solutions 
\cite{Gunaydin:2005mx,Neitzke:2007ke,Gunaydin:2007bg,Gaiotto:2007ag,Cortes:2009cs,Mohaupt:2009iq,Bossard:2009we,Bossard:2009we}, and to study the
radial quantization of BPS solutions \cite{Gunaydin:2005mx,Gunaydin:2007bg}.
The local $c$-map is also an important tool for investigating the 
non-perturbative dynamics of hypermultiplets 
\cite{RoblesLlana:2006ez,RoblesLlana:2006is,Alexandrov:2008gh}, which shows
interesting phenomena such as wall crossing 
\cite{Gaiotto:2008cd,Gaiotto:2009hg}.


The geometry underlying the rigid $r$-map and rigid $c$-map
\cite{Cecotti:1988qn,Cortes:2005uq} is
well understood: for both maps the scalar manifold 
of the higher-dimensional theory is simply replaced by 
its tangent bundle (or, equivalently, its cotangent bundle)
and the special structures on both manifolds are related in a 
canonical way. The metric induced on the (co-)tangent bundle
is a version of the so-called Sasaki metric, where the
special connection rather than the Levi-Civita connection 
is used to define the vertical distribution. To be specific, 
the rigid $r$-map between the scalar manifolds $M,N\simeq TM$
of five- and four-dimensional rigid vector multiplets 
takes the following form in terms of 
adapted coordinates $\sigma^i, b^i$ on $TM$:
\begin{equation}
\label{RigidRmap}
ds^2_M = H_{ij}(\sigma) d \sigma^i d\sigma^j \rightarrow
ds^2_N = H_{ij}(\sigma) (d \sigma^i d\sigma^j + db^i db^j) \;.
\end{equation}

The geometry of the local $r$-map and $c$-map
is more complicated because the supergravity multiplet 
contributes additional degrees of freedom to the scalar manifold.
For the local $r$-map the metric on the 
vector multiplet manifold $\bar{N}$ can nevertheless be brought 
to the above Sasaki form
\cite{Mohaupt:2009iq,Cortes:2011aj,Mohaupt:2011ab}. The reason 
is that the Kaluza-Klein scalar combines with the five-dimensional
scalars precisely in such a way that the scalar manifold $\bar{M}$
of five-dimensional vector multiplets coupled to supergravity is
extended to the scalar manifold $M$ of the associated superconformal
theory, but with the superconformal Hesse potential replaced by 
its logarithm. The scalar manifold $\bar{N}$ of the four-dimensional
vector multiplet theory is then obtained by applying the rigid $r$-map
to $M$.

The local $c$-map \cite{Cecotti:1988qn,Ferrara:1989ik}
has an even more complicated structure. 
It relates projective special K\"ahler manifolds $\bar{N}$ of
dimension $2n$ to quaternion-K\"ahler manifolds $\bar{Q}$ of
dimension $4n+4$. In three dimensions abelian gauge fields,
including the Kaluza-Klein vector can be dualized into scalars,
which become part of the scalar manifold $\bar{Q}$. Using special holomorphic
coordinates on $\bar{N}$, the metric
on $\bar{Q}$ was obtained in \cite{Ferrara:1989ik}. While 
completely explicit, this expression is rather complicated, and
not covariant with respect to the symplectic transformations of the
underlying vector multiplet theory. 

In this paper we reformulate the local $c$-map and obtain
an explicit  expression for the metric in terms of the
Hesse potential of the associated vector multiplet theory
which is symplectically covariant and  only differs from the Sasaki form by 
simple universal terms. This is done using the ideas introduced 
above: (i) we show that the Kaluza-Klein scalar can be identified
with the radial direction of the $\mathbbm{C}^*$-bundle $N$ over
$\bar{N}$. Thus as in the case of the local $r$-map there is
a natural way to combine the four-dimensional scalars with the
Kaluza-Klein scalar. (ii) To preserve symplectic 
covariance we avoid $U(1)$ gauge fixing, which 
amounts to working on a principal $U(1)$ bundle $\hat{Q}$ over
the quaternion-K\"ahler manifold $\bar{Q}$. In complete analogy to 
the vector multiplet case, the metric of $\bar{Q}$ is lifted 
horizontally to a symmetric (degenerate) tensor field on the
total space of $\hat{Q}$. (iii) We use our real formulation
of projective K\"ahler geometry to express everything in terms
of real coordinates and the Hesse potential.

Our construction is different from other `covariant' 
$c$-maps, which use the hyper-K\"ahler cone and twistor
space associated to every quaternion-K\"ahler manifold
\cite{Rocek:2005ij,2006math......3048R,Neitzke:2007ke}. In particular,
the $U(1)$-bundle $\hat{Q}$ and the systematic use of horizontal lifts 
and of special real coordinates are specific to our approach. 
One advantage of our formulation is that we obtain an explicit 
and relatively simple expression for the quaternion-K\"ahler
metric itself. In contrast, other constructions provide expressions
for the hyper-K\"ahler potential of the hyper-K\"ahler cone, or for
the K\"ahler potential of the twistor space, in terms of 
either the holomorphic prepotential \cite{2006math......3048R} 
or the Hesse potential \cite{Neitzke:2007ke}. This leaves
the still complicated step of lifting  
data from 
$\bar{Q}$ to the hyper-K\"ahler cone or twistor space, or projecting
data down from there to $\bar{Q}$. Being able to work directly on 
$\bar{Q}$ has immediate advantages for constructing solutions, as we will 
explain below.

\subsection{Solitons and Instantons}

Dimensional reduction is a standard tool for generating solutions
with (at least) one Killing vector field \cite{Breitenlohner:1987dg}. 
In particular, dimensional
reduction over time allows to lift Euclidean, instantonic solutions
to stationary, solitonic solutions. Therefore we include the case
of time-like reduction when working out the $c$-map. For temporal 
reduction the resulting manifold
has split signature and is expected on general grounds to 
be para-quaternion K\"ahler \cite{Cortes:2003zd,Cortes:2005uq}.

Our main motivation in studying solutions  
is to develop a formalism which does not depend on supersymmetry
(Killing spinor equations), and 
applies to general $c$-map spaces, without the assumption 
that the scalar manifold is symmetric or homogeneous. This
continues work done previously in 
\cite{Cortes:2009cs,Mohaupt:2009iq,Mohaupt:2010du,Mohaupt:2010fk}
for five-dimensional vector multiplets.
For symmetric spaces group theoretical methods have led
to a detailed understanding
of extremal BPS and non-BPS solutions \cite{Gaiotto:2007ag,Bossard:2009we}. 
For general 
$c$-map spaces such methods are not applicable and need to
be replaced by other methods. Solving
the reduced, three-dimensional equations of motion 
is equivalent to finding a harmonic map from the three-dimensional
base space (i.e. the reduced space-time)
into the scalar target space. Particular solutions to this
problem are given by harmonic maps onto totally geodesic
submanifolds \cite{Breitenlohner:1987dg,Cortes:2009cs}.
The simplest choice for the base manifold is
to take it to be flat, which for non-rotating black hole
solutions corresponds to imposing extremality. In this case
the scalar submanifold must be totally isotropic,
so that the classification of BPS and non-BPS non-rotating
solutions corresponds to the classification of totally
geodesic, totally isotropic submanifolds.

In this paper we only consider three-dimensional base spaces which 
are Ricci-flat, and, hence, flat.  We do not impose spherical
symmetry, unless when considering specific examples. One advantage
of our approach is that, for flat base spaces, spherical symmetry
is not needed to solve the field equation, i.e. it is as easy to
obtain multi-centered solutions as single-centered solutions. This is
different in the approach of \cite{Neitzke:2007ke}, where only
single centered BPS black holes were constructed, while multi-centered
solutions were left as an open problem. For this type of problem 
it is advantageous that we 
do not need to lift solutions to the twistor space or to the hyper-K\"ahler 
cone.

The structure of our expression for the (para-)quaternion-K\"ahler metric
immediately suggests that in order to restrict fields to a totally
isotropic submanifold we should 
make an ansatz of the form $\partial_\mu q^a
= \pm \partial_\mu \hat{q}^a$, where the two sets of scalars correspond
to the positive and negative directions of the scalar metric. By lifting
to four dimensions we recognize that this is equivalent to the BPS condition
imposed by the vanishing of the gaugino variation, and we can also
verify that in this case the ADM mass is equal to the central charge.
Thus we have identified totally isotropic 
submanifolds which exist for any $\bar{Q}$ and correspond to BPS
field configurations. As further part of the ansatz we can specify 
whether the solution is rotating or non-rotating. While the non-rotating
solutions include BPS black holes, the rotating solutions are over-extremal,
as expected for rotating BPS solutions in four dimensions. 
By introducing dual coordinates $q_a$ the remaining field equations 
can be brought to the form of decoupled linear 
harmonic equations, $\Delta q_a=0$.
Upon dimensional lifting these equations are recognized as the
black hole attractor equations, which express all fields in terms 
of a set of harmonic functions. 
This is completely analogous to the five-dimensional case. 
To illustrate how the formalism works we include several examples
of rotating and non-rotating solutions. The rotating solutions
we find include those described in 
\cite{Behrndt:1996jn,Bergshoeff:1996gg,Behrndt:1997ny,LopesCardoso:2000qm}. 
For static axion free solutions we show that the solutions previously
known for `very special' prepotentials (those which can be obtained
by dimensional reduction from five dimensions)
can be generalized to a larger
class of prepotentials. The reason is that the ability to solve
the attractor equations only depends on certain homogeneity
properties of the prepotential. A similar observation allowed the
construction of new solutions in five dimensions \cite{Mohaupt:2009iq}.

Extremal non-BPS solutions are associated to totally geodesic, 
totally isotropic submanifolds different from the universal 
ones described above. Since we want to include target spaces 
which are not symmetric, we cannot use the group
theoretical methods of 
\cite{Gaiotto:2007ag,Bossard:2009we} to find non-BPS solutions.  
Another method 
is to replace the central charge by a `fake superportential'
by applying a charge rotation matrix 
\cite{Ceresole:2007wx,LopesCardoso:2007ky}. Within our approach
we can modify the ansatz by allowing a constant field rotation
matrix, $\partial_\mu q^a = R^a_{\;\;b} \partial_\mu \hat{q}^b$, as was done
for the local $r$-map in \cite{Mohaupt:2009iq}. For non-rotating
solutions we show that this generalized ansatz works, but only 
if a compatibility condition between the field rotation matrix
and the metric is satisfied. At first glance this makes it hard
to say anything about the existence of non-BPS extremal solutions
for general, non-symmetric target spaces, without considering
specific models. However, for the class of prepotentials already
mentioned above,
which includes the very special prepotential
as a subclass, we can demonstrate the existence
of a non-trivial field rotation matrix for axion-free solutions. 
In contrast, for rotating solutions the presence of a non-trivial
field rotation matrix always requires to generalize the ansatz
by admitting a curved three-dimensional base space.

\section{Review of  rigid vector multiplets \label{SectionRigidCmap}}

\subsection{Rigid vector multiplets and the rigid $c$-map}

To set the scene, we will review rigid ${\cal N}=2$ vector multiplets and
the rigid $c$-map \cite{Cortes:2003zd,Cortes:2005uq}. We will 
use the conventions of \cite{Cortes:2005uq}, except for a relative minus sign
in the relation between the scalar metric $N_{IJ}$ and 
$\mbox{Im}F_{IJ}$.\footnote{In our present convention the kinetic terms for scalar and vector fields
are positive definite if $\mbox{Im}F_{IJ}$ is positive definite. Note
that if we use the superconformal approach to construct a supergravity theory,
$N_{IJ}$ must be chosen indefinite, with the negative directions corresponding
to conformal compensators.}

Vector multiplets 
$(A^I_{\hat{\mu}}, \lambda^I_i, X^I)$
contain vector fields, a doublet of fermions, and complex scalars.
Here and in the following $\hat{\mu}, \hat{\nu},  \ldots =0,1,2,3$ are Lorentz 
indices, $i=1,2$ is the $SU(2)_R$-index, and $I$ labels the vector
multiplets. The relevant terms in the bosonic Lagrangian are
\begin{eqnarray}
\label{RigidVMlagrangian}
{\cal L}_4 &\sim&  - N_{IJ}(X,\bar{X}) \partial_{\hat{\mu}} 
X^I \partial^{\hat{\mu}}
 \bar{X}^J \\
&& + i \left( F_{IJ}(X) F^{I|-}_{\hat{\mu} \hat{\nu}} F^{J|-|\hat{\mu} 
\hat{\nu}} -
\bar{F}_{IJ}(\bar{X}) F^{I|+}_{\hat{\mu} \hat{\nu} } F^{J|+|\hat{\mu} \hat{\nu}} 
\right) \;, \nonumber
\end{eqnarray}
where $F^{I|\pm}_{\hat{\mu}\hat{\nu}}=\frac{1}{2} \left( F^I_{\hat{\mu}\hat{\nu}}
\mp i \tilde{F}^I_{\hat{\mu}\hat{\nu}}\right)$ are the (anti-)selfdual projections of 
the field strengths $F^I_{\hat{\mu}\hat{\nu}} = 2 \partial_{[\hat{\mu}} 
A^I_{\hat{\nu}]}$. The Hodge-dualization of field strength is given by
$\tilde{F}^I_{\hat{\mu}\hat{\nu}} = \frac{1}{2} \epsilon_{\hat{\mu}\hat{\nu}
\hat{\rho} \hat{\sigma}} F^{I|\hat{\rho} \hat{\sigma}}$. 

All couplings in the Lagrangian can be expressed in terms of
the holomorphic prepotential $F(X^I)$.\footnote{For non-generic choices
of a symplectic frame the prepotential might not exist, but then one
can always perform a symplectic transformation to a frame where a prepotential
exists \cite{Ceresole:1994cx}.} Denoting the derivatives of the prepotential as
\[
F_I = \frac{\partial F}{\partial X^I} \;,\;\;\;
\bar{F}_I = \frac{\partial \bar{F}}{\partial \bar{X}^I} \;,\;\;\;
F_{IJ} = \frac{\partial^2 F}{\partial X^I \partial X^J} \;, \ldots
\]
the scalar metric is 
\[
N_{IJ} = - i (F_{IJ} - \bar{F}_{IJ}) = 2 \mbox{Im}(F_{IJ}) \;.
\]
This is an affine special K\"ahler metric, because the K\"ahler 
potential $K(X,\bar{X})$ for the metric
\[
N_{IJ} = \frac{\partial^2 K}{\partial X^I \partial \bar{X}^J}
\]
can be expressed in terms of the holomorphic prepotential,
\begin{equation}
\label{KaehlerRigid}
K = i (X^I \bar{F}_I - F_I \bar{X}^I) \;.
\end{equation}

The additional, `special' structure of the scalar geometry is 
a consequence of the electric-magnetic duality transformations, which
leave the field equations (but not the action) invariant. Electric-magnetic
duality acts by symplectic transformations, see \cite{deWit:1996ix} for
a concise summary. The quantities
\[
(X^I, F_I)^T \;,\;\;\;(F^{I|\pm}_{\hat{\mu} \hat{\nu}}, G^{\pm}_{I|\hat{\mu}
\hat{\nu}})^T \;,
\]
where the dual gauge fields are defined by
\[
G^{-}_{I|\hat{\mu} \hat{\nu}} = F_{IJ} F^{J|-}_{\hat{\mu}\hat{\nu}} \;,
\]
transform as symplectic vectors, while the second derivatives
$F_{IJ}$ of the prepotential transform fractionally linearly.
The prepotential itself does not transform covariantly, i.e. it is 
not a symplectic function (scalar).

Upon dimensional reduction the components of the gauge fields
along the reduced direction become scalars. After dualizing 
the three-dimensional gauge fields into scalars, one is left
with a theory of scalars and fermions, which organize themselves
into hypermultiplets. The dimensional reductions with respect to 
a space-like and a time-like directions differ by relative signs.
We can discuss both reductions in parallel by introducing the 
parameter $\epsilon$, where $\epsilon=-1$ for space-like and
$\epsilon=+1$ for time-like reductions. We denote scalars 
descending from four-dimensional gauge fields by 
$p^I = A^{I|*}$, where $*=3$ for space-like and $*=0$ for
time-like reductions. The scalars obtained by dualizing the
three-dimensional gauge fields are denoted $s_I$. The scalar
part of the three-dimensional Lagrangian takes the form \cite{Cortes:2005uq} 
\begin{eqnarray} \label{3DrigidLagrangian}
{\cal L}_3 &\sim & -  N_{IJ} \partial_{{\mu}} X^I \partial^{{\mu}} \bar{X}^J 
+ \epsilon  N_{IJ} \partial_{{\mu}} p^I \partial^{{\mu}} p^J \nonumber \\
&& + \epsilon  N^{IJ} (\partial_{{\mu}} s_I + 
R_{IK} \partial_{{\mu}} p^K)(\partial^{{\mu}} s_I + 
R_{JL} \partial^{{\mu}} p^L) \;. 
\end{eqnarray}
Here $N^{IJ}$ is the inverse of $N_{IJ}$, and $\mu, \nu, \ldots =0,1,2$ for 
space-like and $\mu, \nu, \ldots =1,2,3$ for time-like reductions. 

The map induced by dimensional reduction between the
respective scalar manifolds $M$ and $N$ is called the rigid $c$-map.
For space-like reductions $N$ is hyper-K\"ahler \cite{Cecotti:1988qn}, 
as required for rigid 
hypermultiplets \cite{AlvarezGaume:1980dk}. 
For time-like reductions one obtains a
para-hyper-K\"ahler manifold, as required for Euclidean 
hypermultiplets \cite{Cortes:2005uq}.
In both cases the manifold $N$ can be interpreted as the cotangent 
bundle of $M$, $N=T^*M$, equipped with a natural metric,
which one might call the `$\nabla$-Sasaki' metric.\footnote{In contrast
to the Sasaki metric, we use the special connection $\nabla$ instead 
of the Levi-Civita connection to pick a horizontal distribution on $TM$.} 
This becomes
manifest if one uses special real coordinates 
instead of special holomorphic coordinates
on $M$, see (\ref{3dRigidLagrangian}) below. 
Since special real coordinates will play an important role in the
following, we will  review them in some detail. 

\subsection{The real formulation of affine special K\"ahler geometry}

The intrinsic definition of affine special K\"ahler geometry 
\cite{Freed:1997dp} states
that a K\"ahler manifold is affine special K\"ahler\footnote{This 
definition can be generalized
to pseudo-K\"ahler and adapted to para-K\"ahler manifolds 
\cite{Cortes:2003zd}.}
if it is equipped with a flat, torsion-free, symplectic connection $\nabla$,
such that the complex structure $I$ satisfies $d^\nabla I = 0$. 
The affine coordinates $(x^I, y_I)$ of this flat connection are
Darboux coordinates, and are called special real coordinates
in the following. They are related to the special holomorphic 
coordinates $X^I$ by:
\[
x^I = \mbox{Re}(X^I) \;,\;\;\;
y_I = \mbox{Re}(F_I) \;.
\]
Conversely, the special holomorphic coordinates $X^I$ and the 
quantities $F_I$,
which complete them into a complex symplectic vector, can be 
decomposed as
\begin{eqnarray}
X^I &=& x^I + i u^I(x,y) \;,\nonumber \\
F_I &=& y_I + i v_I(x,y) \;. \nonumber
\end{eqnarray}
We remark that we could also take the imaginary 
parts $u^I,v_I$ as real coordinates and $x^I, y_I$
to be functions of $u^I, v_I$. More generally 
we could take the real parts of $e^{i\alpha} (X^I, F_I)$
as Darboux coordinates. Affine special K\"ahler manifolds 
always admit not just one special connection, but an $S^1$-family
which is generated by \cite{Freed:1997dp}
\[
\nabla^{(\alpha)} = e^{\alpha I} \circ \nabla \circ e^{-\alpha I} \;.
\]
Neither physics, nor geometry depends 
on the choice of the special connection from this
family, but each connection in the family has its own system of
special real coordinates. 
The `dual' special real coordinates $u^I, v_I$ are flat
Darboux coordinates with respect to the special connection $\nabla^{(\pi/2)}$.
By  computing the Jacobian of 
the coordinate transformation $(X,\bar{X}) \leftrightarrow (x,y)$,
and using $F_{IJ}=F_{JI}$, 
one obtains the following relations:
\begin{align*}
		\frac{\partial v_I}{\partial x^J} 
&= \frac{\partial v_J}{\partial x^I}\;, & \frac{\partial v_I}{\partial u^J} 
&= \tfrac{1}{2}R_{IJ} \;, \\
		\frac{\partial v_I}{\partial y_J} 
&= -\frac{\partial u^J}{\partial x^I} \;, & \frac{\partial y_I}{\partial u^J} 
&= -\tfrac{1}{2} N_{IJ} \;, \\
		\frac{\partial u^I}{\partial y_J} 
&= \frac{\partial u^J}{\partial y_I} \;,& \frac{\partial u^I}{\partial x^J} 
&= N^{IK}R_{KJ} \;.
\end{align*}
Affine special K\"ahler manifolds are Hessian manifolds, and
the Hesse potential is proportional 
to the Legendre transform of the imaginary part of the
holomorphic prepotential \cite{Cortes:2001}. This transformation 
replaces $u^I = \mbox{Im}X^I$ by 
$y_I = \mbox{Re}F_I$ as independent variables:
\[
H(x,y) = 2 \mbox{Im}F(X(x,y)) - 2 y_I u^I(x,y) \;.
\]
Taking derivatives of $2\, \text{Im}(F)$ with respect to $(x,y)$ we find
	\begin{align*}
		\frac{ \partial }{\partial x^I}  2\text{Im}(F)\Big|_{x,u(x,y)} &= \left(\frac{\partial}{\partial x^I} +  \frac{\partial u^J}{\partial x^I} \frac{ \partial}{\partial u^J}\right) 2\text{Im}(F)\Big|_{x,u} \\
		&= \left[\left(\frac{\partial}{\partial X^I} + \frac{\partial}{\partial \bar{X}{}^I}\right) + i\frac{\partial u^J}{\partial x^I}\left(\frac{\partial}{\partial X^J} - \frac{\partial}{\partial \bar{X}{}^J}\right) \right] 2\text{Im}(F)\Big|_{X,\bar{X}} \\
		&= 2v_I + 2y_I\frac{\partial u^J}{\partial x^I} \;,
	\end{align*}
and
	\begin{align*}
		\frac{ \partial }{\partial y_I}2\text{Im}(F)\Big|_{x,u(x,y)} &= \left(\frac{\partial u^J}{\partial y_I} \frac{ \partial }{\partial u^J}\right) 2\text{Im}(F)\Big|_{x,u} \\
		&= i\frac{\partial u^J}{\partial y_I} \left(\frac{\partial}{\partial X^J} - \frac{\partial}{\partial \bar{X}{}^J} \right)2\text{Im}(F)\Big|_{X,\bar{X}} \\
		&= 2y_J\frac{\partial u^J}{\partial y_I} \;.
	\end{align*}
	Using these results, we find that the derivatives of the 
Hesse potential are proportional to the dual real coordinates:
\begin{equation}\label{Ha}
H_a = \left(\frac{\partial H}{\partial q^a} \right) =
\left(  \frac{\partial H}{\partial x^I} \;, 
\frac{\partial H}{\partial y_I} \right) =
\left( 2v_I \;, -2u^I \right) \;.
\end{equation}
Taking second derivatives we find
	\begin{align*}
		\frac{\partial^2 H}{\partial x^I \partial x^J} &= N_{IJ} + R_{IK}N^{KL}R_{LJ} \;, \\
		\frac{\partial^2 H}{\partial x^I \partial y_J} &= -2N^{IK}R_{KJ}\;, \\
		\frac{\partial^2 H}{\partial y_I \partial y_J} &= 4N^{IJ} \;.
	\end{align*}
This allows us to express the Hessian metric $H_{ab}$ in terms 
of the second derivatives of the prepotential:
\begin{equation}
(H_{ab}) =  \left( \frac{\partial^2 H}{\partial q^a \partial q^b}
\right) = 
\left( \begin{array}{cc}
N + R N^{-1} R & -2 RN^{-1} \\
-2 N^{-1} R & 4 N^{-1} \\
\end{array} \right) \;. \label{eq:H_def}
\end{equation}

We will also need the relation between the differentials
of the special holomorphic and the special real coordinates:
\begin{eqnarray} \label{dX}
dX^M &=& d x^M + i d u^M \nonumber \\
&=& d  x^M + i \left( \frac{\partial u^M}{\partial x^K}
dx^K + \frac{\partial u^M}{\partial y_I }d y_I \right) \nonumber \\
&=& d x^M + i \Big( N^{MI}R_{IK} d x^K  -2N^{MI} 
d y_I \Big) \;.
\end{eqnarray}



Next, we compute the derivatives of the
Hesse potential with respect to the special holomorphic coordinates. This is not needed for the real formulation of affine special K\"ahler geometry, but will be important later for the real formulation of projective affine special K\"ahler geometry. 
	\begin{eqnarray}\label{dHdX}
		\frac{\partial H}{\partial X^I} &=& \frac{\partial x^J}{\partial X^I}\frac{\partial H}{\partial x^J} + \frac{\partial y_J}{\partial X^I}\frac{\partial H}{\partial y_J} \nonumber \\
		&=& 2v_J \frac{\partial x^J}{\partial X^I} - 2u^J  \frac{\partial y_J}{\partial X^I} = v_I - F_{IJ}u^J \nonumber \\
		&=& v_I - \tfrac{1}{2}\left(R_{IJ} + iN_{IJ} \right)u^J \;,
	\end{eqnarray}
	and by a similar calculation
	\[
		\frac{\partial H}{\partial \bar{X}{}^I} = v_I - \tfrac{1}{2}\left(R_{IJ} - iN_{IJ} \right)u^J \;.
	\]
	Taking second derivatives we find 
	\begin{equation}\label{d2HdXdX}
		\frac{\partial^2 H}{\partial \bar{X}{}^J \partial X^I}
= \tfrac{1}{2}N_{IJ} \;.
	\end{equation}


Using equations (\ref{eq:H_def}) and (\ref{dX}) it is straightforward to verify
verify that  
\[
ds^2_M = N_{IJ} d X^I d \bar{X}^J = H_{ab} 
dq^a dq^b \;,
\]
which shows that $N_{IJ}$ and $H_{ab}$ represent the same metric
in terms of special holomorphic and special real coordinates,
respectively. It is easy to show that the
inverse of the Hessian metric is given by
\[
(H^{-1})^{ab} =  (H^{ab}) = \left( \begin{array}{cc}
N^{-1} & \frac{1}{2} N^{-1} R \\
\frac{1}{2} RN^{-1} & \frac{1}{4}(N+RN^{-1}R) \\
\end{array} \right) \;.
\]
Moreover, it is useful to note that 
\[
H_{ab} \Omega^{bc} H_{cd} = - 4 \Omega_{ad} \;,
\]
where 
	\[
		\Omega_{ab} := \left( \begin{array}{cc} 0 &	\mathbb{I}	\\ -\mathbb{I} & 0 \end{array} \right) 
	\]
is the matrix representing the fundamental form (K\"ahler form) in 
special real coordinates.\footnote{The fundamental form has constant
coefficients because special real coordinates are Darboux coordinates.} 
With these results it is straightforward to express the reduced Lagrangian
(\ref{3DrigidLagrangian}) in terms of special real coordinates.
Defining $(\hat{q}_a) = (s_I, 2 p^I)$, we find 
\cite{Cortes:2005uq,Mohaupt:2007md}
\begin{equation}
\label{3dRigidLagrangian}
{\cal L}_3 \sim - \left( H_{ab}(q) \partial_\mu q^a \partial^\mu q^b 
-\epsilon H^{ab}(q) \partial_\mu \hat{q}_a \partial^\mu \hat{q}_b \right) \;.
\end{equation}
It is now manifest that the metric on $N$ is the canonical 
positive definite (for $\epsilon=-1$) and split signature (for $\epsilon =1$)
metric 
on the cotangent bundle of $M$, respectively. Using special real coordinates
has further advantages. All objects appearing in the above Lagrangian
transform linearly under symplectic transformations: $q^a$, $\hat{q}_a$
are contravariant and covariant vectors, respectively, while $H_{ab}$ and
$H^{ab}$ are symmetric tensors \cite{Cortes:2011aj}. 
In contrast, $F_{IJ}=\frac{1}{2} (R_{IJ} 
+ i N_{IJ})$ transforms fractionally linearly under symplectic transformations.

\section{Vector multiplets coupled to 4d 
supergravity \label{Sec:LocalVM}}

The coupling of vector multiplets to supergravity can be constructed
using the superconformal calculus, which exploits the gauge 
equivalence between a locally superconformal theory of $n+1$ 
vector multiplets and $n$ vector multiplets coupled to Poincar\'e
supergravity.\footnote{This requires the presence of a further 
auxiliary multiplet, which will not be relevant for our discussion.}
This is reviewed, for example, in \cite{Mohaupt:2000mj,deWit:2002vz}.
We will use elements of this approach, and focus on the bosonic
fields and the underlying scalar geometry. The first step in 
the construction is to write down a theory of $n+1$ rigidly
superconformal vector multiplets. Compared to the previous
section, this amounts to the additional constraint that the
prepotential is homogeneous of degree two. The resulting
scalar manifold is a conical affine special K\"ahler manifold 
\cite{Freed:1997dp,Cortes:2009cs},
which is an affine special K\"ahler manifold with a holomorphic 
homothetic action of $\mathbbm{C}^* = \mathbbm{R}^{>0}\cdot U(1)$:
\[
X^I \rightarrow \lambda X^I \;,
\]
where $\lambda = |\lambda|e^{i\phi} \in \mathbbm{C}^*$. Both the
scale transformation and the $U(1)$ phase transformation are part 
of the superconformal algebra. The scale transformations act as
homotheties, and give the scalar manifold $N$ the structure of a 
Riemannian cone over a Sasakian manifold $S$. The $U(1)$ transformations
act isometrically on both $N$ and $S$. 

The next step in the superconformal construction is to gauge the
superconformal transformations. For our purposes, the relevant
part of the resulting bosonic action is
\begin{equation}
\label{LocalVMlagrangian}
		{\cal L}_4 \sim -\tfrac{1}{2} e^{-{\cal K}} R_4 - N_{I J} {\cal D}_{\hat{\mu}} X^I {\cal D}^{\hat{\mu}} \bar{X}{}^J  + \tfrac{1}{4} {\cal I}_{IJ} F^I_{\hat{\mu} \hat{\nu}} F^{J \hat{\mu} \hat{\nu}} + \tfrac{1}{4} {\cal R}_{IJ} F^I_{\hat{\mu} \hat{\nu}} \tilde{F}^{J \hat{\mu} \hat{\nu}} \;,
\end{equation}
	where the indices run from $I = 0,\ldots,n$.
This Lagrangian contains the space-time Ricci scalar $R_4$ as a result
of the gauging. It is invariant under local dilatations 
and $U(1)$ dilatations. The $U(1)$ covariant derivatives are defined by
	\begin{align*}
		{\cal D}_{\hat{\mu}} X^I &= (\partial_{\hat{\mu}} + iA_{\hat{\mu}}) X^I \;, \\
		{\cal D}_{\hat{\mu}} \bar{X}{}^I &= (\partial_{\hat{\mu}} - iA_{\hat{\mu}}) \bar{X}{}^I	\;,
	\end{align*}
where $A_{\hat{\mu}}$ is the $U(1)$ connection. In principle we should also 
include the connection $b_{\hat{\mu}}$ of local dilatations, but it is known 
that the terms containing this connection cancel within the Lagrangian. 
Alternatively, one can 
impose the gauge condition $b_{\hat{\mu}}=0$, known as the K-gauge. 
The gravitational term is not canonical, since the Ricci scalar is 
multiplied by the dependent field 
	\begin{equation}
\label{DilatationalCompensator}
		e^{-{\cal K}} = -N_{IJ}X^I \bar{X}{}^J =
-i (X^I \bar{F}_I - F_I \bar{X}^I) \;,
	\end{equation}
which acts as a compensator for local dilatations.

The gauge couplings are given by the real and imaginary parts of the 
complex matrix
	\begin{equation}
\label{DefCalN}
		{\cal N}_{IJ} = {\cal R}_{IJ} + i {\cal I}_{IJ} = \bar{F}_{IJ} + i \dfrac{(NX)_I (NX)_J}{XNX} \;.
	\end{equation}
This differs from the gauge couplings $F_{IJ} = \frac{1}{2}(R_{IJ} + i N_{IJ})$
of the rigid theory
by terms which arise from integrating out an auxiliary field (the 
tensor field of the Weyl multiplet). Note that ${\cal N}_{IJ}$
is manifestly $U(1)$ invariant, so that by imposing the $D$-gauge
we obtain tensor fields on $S$ and $\bar{N}$.

The locally superconformal Lagrangian, of which we have displayed only
the pieces relevant for our purposes, is gauge equivalent to a Lagrangian
of vector multiplets coupled to Poincar\'e supergravity. The Poincar\'e
supergravity Lagrangian is obtained by imposing conditions which gauge
fix the additional transformations which extend the Poincar\'e supersymmetry
algebra to the superconformal algebra. For our purposes the relevant 
transformations are the dilatations and $U(1)$ transformations. 
The dilatations are gauge fixed by imposing the D-gauge
$e^{-\cal K} =1$, which brings the gravitational term to its canonical,
Einstein-Hilbert form. Geometrically, this restricts the scalar fields
to a hypersurface ${\cal H} \subset N$ in the conical affine special
K\"ahler manifold. This hypersurface can be identified with the 
Sasakian $S$, which forms the basis of the Riemannian cone. Similarly,
one can impose a $U(1)$ gauge condition to obtain the scalar manifolds
$\bar{N}$ of the Poincar\'e supergravity theory. In practice, one often
prefers to work in terms of $U(1)$ invariant quantities instead of imposing
an explicit gauge fixing condition. 
Since the $U(1)$ transformations act isometrically on $S$, this corresponds
to taking a quotient $S/U(1)$. Moreover, since the function $e^{-{\cal K}}$
used to define the D-gauge is the moment map of the $U(1)$ isometry,
the scalar manifolds $N$ and $\bar{N}$ of the superconformal and 
super-Poincar\'e theories are related by a symplectic quotient
\[
\bar{N} \simeq N / \mathbbm{C}^* \simeq N // U(1) \;.
\]
This is in fact a K\"ahler quotient, because $\bar{N}$ inherits
a K\"ahler metric from $N$. Manifolds $\bar{N}$, which are obtained
by this construction from conical affine special K\"ahler manifolds,
are called projective special K\"ahler manifolds. 

It is well known from work on black hole solutions that it is
often advantageous to use the gauge equivalence, and to work on the
larger space $N$ rather than on the physical scalar manifold 
$\bar{N}$. One particular advantage is that this keeps symplectic
covariance manifest. Fixing a $U(1)$ gauge corresponds to selecting
a hypersurface of the Sasakian $S$, which can be done by choosing
any condition which is transversal to the $U(1)$ action (for example
$\mbox{Im} X^0=0$). However, choosing a symplectically invariant
condition corresponds to selecting, at each point, the direction 
orthogonal to the $U(1)$ action. But this is the contact distribution 
of the Sasakian and therefore not integrable. For this reason a hypersurface
corresponding to a $U(1)$ gauge cannot be selected in a symplectically
invariant way.\footnote{We thank Vicente Cort\'es for an illuminating
discussion of this point.} In the following we will keep the
local $U(1)$ gauge invariance intact, and for reasons that will 
become clear later we also postpone  imposing the D-gauge.

The above Lagrangian contains the $U(1)$ gauge field, which makes
its local $U(1)$ invariance manifest. However, the $U(1)$ connection
is a non-dynamical, auxiliary field, and we now 
eliminate it by its equation of motion
	\[
		A_{\hat{\mu}} = -\tfrac{i}{2} e^{\cal K}\left[ (\partial_{\hat{\mu}} X)N\bar{X} - XN(\partial_{\hat{\mu}}\bar{X}) \right]\;.
	\]
Now the gauged sigma model is replaced by the ungauged sigma model
	\begin{align*}
		- N_{I \bar{J}} {\cal D}_{\hat{\mu}} X^I {\cal D}^{\hat{\mu}} \bar{X}{}^J &= -\left(N_{IJ} - \frac{(NX)_I (N\bar{X})_J}{XN\bar{X}} \right) \partial_{\hat{\mu}} X^I \partial^{\hat{\mu}} \bar{X}{}^J  + \tfrac{1}{4} e^{-{\cal K}} \partial_{\hat{\mu}} {\cal K} \partial^{\hat{\mu}} {\cal K} \;, \\
		&= - e^{-{\cal K}} g_{IJ} \partial_{\hat{\mu}} X^I \partial^{\hat{\mu}} \bar{X}{}^J  + \tfrac{1}{4} e^{-{\cal K}} \partial_{\hat{\mu}} {\cal K} \partial^{\hat{\mu}} {\cal K} \;,
	\end{align*}
	where $g_{IJ} = \partial_I \partial_{\bar{J}} {\cal K}$,
\begin{equation}
\label{KaehlerLocalUp}
{\cal K} = - \log [-i(X^I \bar{F}_I - F_I \bar{X}^I)] \;.
\end{equation}
We have used that the prepotential is homogeneous of degree 2 and therefore $X(\partial_{\hat{\mu}} N) \bar{X} = 0$. 
The Lagrangian still contains terms proportional to 
$\partial_{\hat{\mu}}{\cal K}$ because we 
have not yet imposed the D-gauge. Observe that the tensor field
$g_{IJ}$ is degenerate on the large space $N$, because
\[
X^I g_{IJ} = 0 = g_{IJ} \bar{X}^J \;.
\]
This is not a problem, because the directions along which
$g_{IJ}$ is degenerate correspond to the unphysical degrees of 
freedom normal to $\bar{N} \subset N$. Geometrically, these are
the vertical directions of the $\mathbbm{C}^*$-bundle $N$ over $\bar{N}$,
 i.e. the radial direction of the
Riemannian cone and the orbits of the $U(1)$ isometry. While $g_{IJ}$
is not a metric on $N$, we obtain a non-degenerate metric by projecting
it $\bar{N}$. In other words, $g_{IJ}$ is the horizontal lift 
of the projective special K\"ahler metric $g_{\bar{N}}$ to $N$, and,
if we impose the D-gauge, to $S$.

The well known formula for the K\"ahler potential of the projective
special K\"ahler manifold $\bar{N}$ can be obtained  
by using coordinates $X^0, z^i$ on $N$, where 
$z^i = X^i/X^0$ are special coordinates on $\bar{N}$. Rewriting ${\cal K}$
given in (\ref{KaehlerLocalUp})
as a function of $X^0, z^i$, one finds that the dependence on $X^0$
can be removed by a K\"ahler transformation. This shows explicitly that
the tensor $g_{IJ}$ is degenerate on the two vertical directions.
Defining 
${\cal F}(z)  = (X^0)^{-2} F(X^I)$, we obtain the K\"ahler 
potential of the projective special K\"ahler metric of $\bar{N}$:
\[
{\cal K} = -\log(-i[ ({\cal F} - \bar{\cal F}) - (z^i -\bar{z}^i)
({\cal F}_i + \bar{\cal F}_i)]) \;,\;\;\;
{\cal F}_i = \frac{\partial {\cal F}}{\partial z^i} \;.
\]

To obtain a theory with positive definite kinetic terms for the
physical scalars, the projection of $g_{IJ}$ onto $\bar{N}$ must be
positive definite, while positive definite kinetic terms for 
the vector fields require
that ${\cal I}_{IJ}$ is negative definite, see (\ref{LocalVMlagrangian}).
It is known that both conditions are satisfied if the metric 
$N_{IJ}$ of the conical affine special K\"ahler manifold $N$ has
complex Lorentz signature $(--+\cdots +)$ \cite{Cremmer:1984hj,Cortes:2011aj}. 
The negative directions,
which are the directions normal to $\bar{N}\subset N$, correspond
to conformal compensators. We remark that $-{\cal I}_{IJ}$ can
be interpreted as a positive definite metric on $N$, and that
the relation between the indefinite metric $N_{IJ}$ and the definite
metric $-{\cal I}_{IJ}$
has a natural geometric interpretation, which is analogous to the
relation between the Griffith and Weil intermediate Jacobians
for Calabi-Yau threefolds \cite{Cortes:2011aj}.

\section{The real formulation of projective special K\"ahler 
geometry \label{Sect:Real}}

In section \ref{SectionRigidCmap} we have reviewed the real
formulation of affine special K\"ahler geometry. 
It is not straightforward to obtain a real formulation of
projective special K\"ahler geometry which preserves symplectic
covariance. The reason is that
the physical scalars of the super-Poincar\'e theory correspond
to special coordinates $z^i = \frac{X^i}{X^0}$ on $\bar{N}$. While 
$(X^I, F_I)$ is a symplectic vector, the $(z^i)$ is not, and
only part of the symplectic covariance can be kept 
manifest \cite{Ferrara:2006at}.

In this section we show how a manifestly symplectic real formulation
can be obtained by preserving the $U(1)$ gauge invariance. This amounts
to expressing the degenerate tensor $g_{IJ}$ and the 
vector kinetic matrix ${\cal N}_{IJ}$ in terms of special real coordinates
on $N$ and in terms of the Hesse potential $H$. In doing so we will
get a clearer understanding of the geometrical meaning of these tensor
fields.

Since the theory associated with $N$ is now superconformal,
we have additional relations in addition to those derived
in section \ref{SectionRigidCmap}. The prepotential and the
Hesse potential are now homogeneous of degree two in 
special holomorphic and special real coordinates, respectively.
This implies
\begin{equation}
2H = H_a q^a = H_{ab} q^a q^b \;. \label{eq:H_hom_props}
\end{equation}
Also note that 
\begin{equation}
\label{eq:phi_K}
2(y_I u^I - x^I v_I) = - 2 H = 
-i (X^I \bar{F}_I - F_I \bar{X}^I) = - N_{IJ} X^I \bar{X}^J = e^{-{\cal K}}\;.
\end{equation}

The affine special K\"ahler manifold is now a complex cone,
at least locally. This means that there is a homothetic and
holomorphic action of $\mathbbm{C}^*$, which is given by
the the homothetic Killing vector 
field $\xi$ and the $U(1)$ Killing vector field $I\xi$, where
$I$ is the complex structure. The explicit expressions
with respect to special holomorphic and special real 
coordinates are:
\begin{eqnarray}
\xi &=& X^I \frac{\partial}{\partial X^I} + 
\bar{X}^I \frac{\partial}{\partial \bar{X}^I}  = 
q^a \frac{\partial}{\partial q^a} \;, \nonumber \\
I\xi &=&   i X^I \frac{\partial}{\partial X^I} -i 
\bar{X}^I \frac{\partial}{\partial \bar{X}^I}  =
\tfrac{1}{2} H_a \Omega^{ab} \frac{\partial}{\partial q^b} \;.
\nonumber
\end{eqnarray}
In special real coordinates 
the complex structure itself is given by 
$I^a_{\;\;c} = \frac{1}{2}\Omega^{ab} H_{bc}$ in terms of 
the K\"ahler form $\Omega_{ab}$ and the metric $H_{ab}$. 

We remark that the $q^a$ are special real coordinates with respect
to a fixed, but arbitrary special connection. For conical affine
special K\"ahler manifolds the $U(1)$ gauge transformations preserve
the metric, the symplectic and the complex structure, but they 
rotate the special connections, 
and the associated special real coordinates, among
themselves. 

Our first task is to rewrite the tensor 
\begin{equation}
		g_{IJ}  = \frac{\partial^2 \,{\cal K}}{\partial X^I \partial 
\bar{X}{}^J} = -\frac{Y_{I\bar{J}}}{Y} + \frac{Y_I Y_{\bar{J}}}{Y^2}
\;,
\end{equation}
where
	\[
{\cal K} =-\log Y \;,\;\;\;	Y= -i (X^I\bar{F}_I - F_I \bar{X}^I) 
=-2H \;,\
	\]
in terms of special real coordinates.
Using (\ref{dHdX}) and (\ref{d2HdXdX}) we find	
\begin{equation}\label{K}
		g_{IJ} = -\frac{1}{2H}N_{IJ} + \frac{1}{H^2}\left(v_I - \tfrac{1}{2}\left(R_{IK} + iN_{IK} \right)u^K \right)\left(v_J - \tfrac{1}{2}\left(R_{JL} - iN_{JL} \right)u^L \right) \;.
\end{equation}
Using (\ref{K}), we find
\begin{eqnarray}\label{KXX}
&& K_{I\bar{J}} dX^I d\bar{X}^J = - \frac{1}{2H} N_{IJ} dX^I d\bar{X}^J \\
&&
+ \frac{1}{H^2} (v_I - \frac{1}{2} (R_{IK} + i N_{IK})u^K)
(v_J - \frac{1}{2} (R_{JL} - i N_{JL})u^L) dX^I d\bar{X}^J \;.
\nonumber \end{eqnarray}
By the results of section \ref{SectionRigidCmap}, the first term
gives
\[
- \frac{1}{2H} N_{IJ} dX^I d\bar{X}^J = - \frac{1}{2H} H_{ab}
dq^a dq^b \;.
\]
To evaluate the second term, we observe that 
\[
(2v_I, -2u^I) =
(H_a) = (H_{ab} q^b) \;,
\]
where we used (\ref{Ha}) together with homogeneity. 
Using further results from section \ref{SectionRigidCmap},
this implies
\begin{eqnarray}
x^I &=& 2N^{IJ} v_J - N^{IJ} R_{JK} u^I \;, \nonumber \\
y_I &=& R_{IJ} N^{JK} v_K - \frac{1}{2} (N_{IJ} + R_{IK} N^{KL} R_{LJ}) u^J \;.
\nonumber 
\end{eqnarray}
To proceed, we substitute (\ref{dX}) into the second term on the
right hand side of (\ref{KXX}), with the result
\begin{eqnarray}
&&
(v_I - \frac{1}{2} (R_{IK} + i N_{IK})u^K)
(v_J - \frac{1}{2} (R_{JL} - i N_{JL})u^L) dX^I d\bar{X}^J 
\nonumber \\
&=& (v_I v_J + y_I y_J) dx^I dx^J - (v_I u^J + y_I x^J) dx^I dy_J \nonumber \\
&& - (u^I v_J + x^I y_J) dy_I dx^J + (u^I u^J + x^I x^J) dy_I dy_J  \;.
\nonumber
\end{eqnarray}

We now observe that
\[
(H_a H_b) = 4 \left( \begin{array}{cc}
v_I v_J & - v_I u^J \\
-u^I v_J & u^I u^J \\
\end{array} \right) 
\]
and
\[
(\Omega_{ac}q^c \Omega_{bd} q^d) = \left( \begin{array}{cc}
y_I y_J & - y_I x^J \\
- x^I y_J & x^I x^J \\
\end{array} \right) \;.
\]
Using this, the second term becomes
\begin{eqnarray}
&&
\frac{1}{H^2} \Big( (v_I v_J + y_I y_J) dx^I dx^J - (v_I u^J + y_I x^J) dx^I dy_J \nonumber \\
&& - (u^I v_J + x^I y_J) dy_I dx^J + (u^I u^J + x^I x^J) dy_I dy_J \Big) 
\nonumber \\
&=& 
\left(\frac{1}{4H^2} H_a H_b + \frac{1}{H^2} (\Omega_{ac} q^c \Omega_{bd} 
q^d) \right)  dq^a dq^b\;. \nonumber 
\end{eqnarray}
Combining the two terms, we find that
\begin{equation}
\label{DefH0}
 g_{IJ} d X^I d \bar{X}^J =
\left[ -\frac{1}{2H} H_{ab} + \frac{1}{4H^2}H_a H_b 
+ \frac{1}{H^2} (\Omega_{ac} q^c \Omega_{bd} q^d) 
\right] d q^a d q^b =: H^{(0)}_{ab} dq^a dq^b\;,
\end{equation}
where $H^{(0)}_{ab}$ is the horizontal lift of the 
projective special K\"ahler metric, expressed in special real
coordinates.

Before we proceed to express ${\cal N}_{IJ}$ in real coordinates,
let us analyze what the above calculation tells us about
the underlying geometry. Solving (\ref{DefH0}) for the affine
special K\"ahler metric $H_{ab}$, we obtain:
\[
H_{ab} = - 2H H^{(0)}_{ab} + \frac{1}{2H} H_a H_b + \frac{2}{H}
\Omega_{ac}q^c \Omega_{bd} q^d 
\]
This is a decomposition of $H_{ab}$ into the horizontal 
component $H^{(0)}_{ab}$, which by projection gives the
projective special K\"ahler metric, and two negative
definite terms\footnote{With our conventions $H$ is negative
definite, see (\ref{eq:phi_K}).} which correspond to the directions generated
by $\xi = q^a \partial_a$ and $I\xi = \frac{1}{2} H_a \Omega^{ab}
\partial_b$. As we will see, all relevant tensor fields on $N$ are
related to the metric $H_{ab}$ by adding terms proportional to the
squares of the one-forms $H_a$ and $\Omega_{ac}q^c$. These one forms
are obtained by contracting the homothety $\xi$ with the metric
and the K\"ahler form, respectively (equivalently by contracting
$\xi$ and $I\xi$ with the metric). It is an advantage of the 
real formalism that the directions generated by $\xi$ and $I\xi$ can
be described in such a simple way.

We now introduce one further tensor field on $N$, which will
play an important role for the $c$-map. 
As we have seen before, the K\"ahler potential (\ref{KaehlerLocalUp})
of the supergravity theory is obtained by taking the logarithm of 
the K\"ahler potential (\ref{KaehlerRigid}) of the corresponding
superconformal theory.  The logarithm effectively encodes the
superconformal quotient.  This motivates us introduce the 
tensor obtained by taking the second derivatives of the
logarithm of the Hesse potential $H$ of
the rigid theory. Specifically, we set $\tilde{H} = -\frac{1}{2}\log (-2H)$
and $\tilde{H}_{ab} = \partial^2_{a,b} \tilde{H}$. 
Then
	\begin{equation}
		g_{IJ} dX^I d\bar{X}^J = \left[ \tilde{H}_{ab} - \frac{1}{4H^2}H_a H_b + \frac{1}{H^2}(\Omega_{ac}q^c)(\Omega_{bd}q^d) \right] d q^a d q^b \;.
		\label{eq:real_decomp}
	\end{equation}
Since we know that the right hand side is positive definite in the
horizontal directions and degenerate in the vertical directions generated
by $\xi$ and $I\xi$, it follows immediately that $\tilde{H}_{ab}$ is a 
non-degenerate Hessian metric which is negative definite along 
the $U(1)$ direction generated by $I\xi$ and positive definite in 
all other directions. 
The homogeneity properties of the Hesse potential (\ref{eq:H_hom_props})
also imply that the matrix $\tilde{H}$ satisfies the identity
	\[
		q^a q^b \tilde{H}_{ab} = 1 \;. 
	\]
	This will not be used in this paper, but may be useful for produce 4d non-extremal black hole solutions as a similar identity was needed in the 5d case \cite{Mohaupt:2010fk}.

We now turn to the vector kinetic matrix ${\cal N}_{IJ}$. 
It is known how to express this complex matrix, which transforms
fractionally linearly under symplectic transformations in 
terms of a real matrix $\hat{H}_{ab}$, which transforms as a symmetric
tensor of rank 2. In the conventions of \cite{Cortes:2011aj},
the relation is
		\[
		\hat{H}_{ab} :=  \left( \begin{array}{cc} {\cal I} + {\cal R}{\cal I}^{-1} {\cal R} &	- {\cal R} {\cal I}^{-1}	\\ \vspace{-1.5ex} \\ -{\cal I}^{-1} {\cal R} & {\cal I}^{-1} \end{array} \right) \;.
	\]
It is known that the tensor $-\hat{H}_{ab}$ is positive definite,
given that $H_{ab}$ has complex Lorentz structure, and therefore
it can be interpreted as a positive definite metric on $N$. 
In \cite{Cortes:2011aj} it was shown that in terms of complex 
geometry the indefinite and definite metric are related by 
a transformation that exchanges Griffith and Weyl flags. We would
now like to relate $\hat{H}_{ab}$ to the other tensor fields
in terms of real coordinates.

Below we will prove that the tensors $\tilde{H}_{ab}$, $H_{ab}$ and
$\hat{H}_{ab}$ are related by:
	\begin{align}
		\tilde{H}_{ab} &= -\frac{1}{2H} H_{ab} + \frac{1}{2H^2}H_a H_b \notag \\
									 &= \frac{1}{H} \hat{H}_{ab} - \frac{2}{H^2}(\Omega_{ac}q^c)(\Omega_{bd}q^d) \;. \label{eq:Hess_M_identity}
	\end{align} 
Given that $H_{ab}$ has complex Lorentz signature, it is manifest that
$-\hat{H}_{ab}$ is positive definite. In contrast to the indefinite
metrics $H_{ab}$ and $\tilde{H}_{ab}$, the definite metric $\hat{H}_{ab}$
is not Hessian. However it is uniquely determined by the Hesse potential
$H$. It is the above  identity which will be critical in matching up moduli 
fields with the electric/magnetic potentials in order to produce black hole 
solutions.

It remains to prove (\ref{eq:Hess_M_identity}), which requires some
effort. To start we need the explicit relations between
the real and imaginary parts of $F_{IJ}=\frac{1}{2}(R_{IJ} + i N_{IJ})$
and ${\cal N}_{IJ}= {\cal R}_{IJ} + i {\cal I}_{IJ}$:
\begin{eqnarray}
{\cal R}_{IJ} &=& \frac{1}{2} R_{IJ} + \frac{i}{2}
\left( \frac{ N_{IK}X^K N_{JL} X^L }{(XNX)} -
\frac{N_{IK}\bar{X}^K N_{JL} \bar{X}^L }{(\bar{X}N\bar{X})} \right)
\nonumber \\
{\cal I}_{IJ} &=& - \frac{1}{2} N_{IJ} + \frac{1}{2}
\left( \frac{ N_{IK}X^K N_{JL} X^L }{(XNX)} +
\frac{N_{IK}\bar{X}^K N_{JL} \bar{X}^L }{(\bar{X}N\bar{X})} \right)
\;, \nonumber
\end{eqnarray}
where $(XNX) = N_{MN}X^M X^N$, etc. It is straightforward to verify that
the inverse of ${\cal I}_{IJ}$ is
\[
{\cal I}^{IJ} = -2 N^{IJ} + \frac{2}{(XN\bar{X})} \left( X^I \bar{X}^J
+ \bar{X}^I X^J \right)  = -2 N^{IJ} + \frac{2}{H} \left( x^I x^J +
u^I u^J \right)\;,
\]
where we used $2H = (XN\bar{X})$ and the decomposition 
$X^I = x^I + i u^I$, $F_I = y_I + i v_I$. Next, one can verify
\[
- {\cal I}^{IK} {\cal R}_{KJ} = N^{IK} R_{KJ} - \frac{2}{(XN\bar{X})}
(X^I \bar{F}_J + \bar{X}^I F_J) = N^{IK} R_{KJ} - \frac{2}{H}
(x^I y_J + u^I v_J ) \;.
\]
Finally, one can also verify that
\begin{eqnarray}
&&{\cal I}_{IJ} + {\cal R}_{IK} {\cal I}^{KL} {\cal R}_{LJ} =
- \frac{1}{2} N_{IJ} - \frac{1}{2} R_{IK} N^{KL} R_{LJ} 
+ \frac{2}{(XN\bar{X})} ( F_I \bar{F}_J + \bar{F}_I F_J)  \nonumber \\
&=&  - \frac{1}{2} N_{IJ} - \frac{1}{2} R_{IK} N^{KL} R_{LJ}  + \frac{2}{H}
(y_I y_J + v_I v_J ) \nonumber  
\end{eqnarray}
Putting everything together we have 
\begin{eqnarray}
\left( \begin{array}{cc}
{\cal I} + {\cal R} {\cal I}^{-1} {\cal R} & - {\cal R} {\cal I}^{-1} \\
- {\cal I}^{-1} {\cal R} & {\cal I}^{-1} \\
\end{array} \right) &=& 
\left( \begin{array}{cc} 
- \frac{1}{2} N -\frac{1}{2} RN^{-1}R & RN^{-1} \\
RN^{-1} & - 2 N^{-1} \\
\end{array} \right) \nonumber \\
&& + \frac{2}{H} \left( \begin{array}{cc}
y_I y_J + v_I v_J & -(y_I x^J + v_I u^J ) \\
- (x^I y_J  + u^I v_J) & x^I x^J + u^I u^J \\
\end{array} \right) \;.\nonumber 
\end{eqnarray}
Expressing this in terms of 
the special real coordinates $q^a$ using 
$\hat{H}_{ab}$, $H_{ab}$ and $\Omega_{ab}$ this
becomes
\[
\hat{H}_{ab} = - \frac{1}{2} H_{ab} + \frac{2}{H} 
\left( \frac{1}{4} H_{a}H_{b} + \Omega_{ac} q^c \Omega_{bd} q^d
\right) \;.
\]
which proves (\ref{eq:Hess_M_identity}). 


In summary, we have found the real 
tensor fields $H^{(0)}_{ab}$ and $\hat{H}_{ab}$ which lift
the scalar metric and vector kinetic matrix of the 
super-Poincar\'e theory associated to $\bar{N}$ to 
the Sasakian $S$ and the complex cone $N$. This provides
a real formulation of projective special K\"ahler geometry
as long as we do not gauge fix the $U(1)$ transformations.

For later use we now derive the expression for the graviphoton in terms
of real coordinates. The graviphoton is the vector field which 
in the Poincar\'e supergravity theory belongs to the supergravity
multiplet and therefore is invariant under symplectic transformations. 
Its field strength is given by
\[
T^-_{\hat{\mu}\hat{\nu}} = - X^I G^{-}_{I|\hat{\mu}\hat{\nu}} 
+ F_I F^{I|-}_{\hat{\mu}\hat{\nu}}  \;,
\]
where the dual field strength are 
\[
G^-_{\hat{\mu}\hat{\nu}} = \bar{\cal N}_{IJ} F^{J|-}_{\hat{\mu}\hat{\nu}} \;.
\]
Adding the self-dual part and expressing everything in real variables,
we obtain
\[
T_{\hat{\mu}\hat{\nu}} = T^+_{\hat{\mu}\hat{\nu}} +
T^-_{\hat{\mu}\hat{\nu}} = 
- x^I G_{I|\hat{\mu}\hat{\nu}} + y_I F^I_{\hat{\mu}\hat{\nu}} 
+ u^I \tilde{G}_{I|\hat{\mu}\hat{\nu}} - v_I \tilde{F}^I_{\hat{\mu}
\hat{\nu}} \;.
\]
These terms are not independent, we can either use the real coordinates
$(x^I,y_I)$ together with the field strength $F^I_{\hat{\mu}\hat{\nu}},
G_{I|\hat{\mu}\hat{\nu}}$, or the dual real coordinates $(u^I,v_I)$
together with the Hodge-dual field strength 
$\tilde{F}^I_{\hat{\mu}\hat{\nu}}, \tilde{G}_{I|\hat{\mu}\hat{\nu}}$.
Using the definitions of these quantities, one can verify that
\[
x^I G_{I|\hat{\mu}\hat{\nu}} - y_I F^I_{\hat{\mu} \hat{\nu}} =
v_I \tilde{F}^I_{\hat{\mu} \hat{\nu}} - u^I \tilde{G}_{I|\hat{\mu}
\hat{\nu}} \;,
\]
so that
\[
T_{\hat{\mu}\hat{\nu}} = 
- 2 \left( x^I G_{I|\hat{\mu}\hat{\nu}} - y_I F^I_{\hat{\mu}\hat{\nu}} \right) = 
2 \left( u^I \tilde{G}_{I|\hat{\mu}\hat{\nu}} - v_I \tilde{F}^I_{\hat{\mu}
\hat{\nu}} \right) \;.
\]

\section{The local $c$-map and the Hesse potential}

We now turn to the dimensional reduction of four-dimensional
vector multiplets coupled to supergravity. We perform the 
reduction using the complex formulation of the four-dimensional
scalars, and use the gauge equivalence to describe them in 
terms of the scalars $X^I$ taking values in $N$. The reductions
over space and time are performed in parallel.
After dualizing the three-dimensional vector fields 
we systematically express all quantities in terms of special
real coordinates. Our overall strategy is to obtain an expression 
which comes as close to the `metric on the cotangent bundle form'
(\ref{3dRigidLagrangian})
of the rigid $c$-map. Therefore we express the couplings
in terms of the logarithm of the Hesse potential.
We will see that all terms that cannot be brought to this 
form are universal, in the sense that 
their couplings only contain constant matrices and
the Kaluzu-Klein scalar.

\subsection{Dimensional reduction}

	Our starting point is the Lagrangian representing the bosonic part of four-dimensional ${\cal N} = 2$ supergravity coupled to $n$ vector 
multiplets,
\begin{eqnarray}
		{\cal L}_4 &\sim& -\tfrac{1}{2} e^{-{\cal K}} R_4 
- e^{-{\cal K}} g_{IJ} \partial_{\hat{\mu}} X^I \partial^{\hat{\mu}} \bar{X}{}^J  + \tfrac{1}{4} e^{-{\cal K}} \partial_{\hat{\mu}} {\cal K} \partial^{\hat{\mu}} {\cal K} \nonumber \\
&& + \tfrac{1}{4} {\cal I}_{IJ} F^I_{\hat{\mu} \hat{\nu}} F^{J \hat{\mu} \hat{\nu}} + \tfrac{1}{4} {\cal R}_{IJ} F^I_{\hat{\mu} \hat{\nu}} \tilde{F}^{J \hat{\mu} \hat{\nu}} \;,
\end{eqnarray}
where $g_{IJ} = \partial_I \partial_{\bar{J}} {\cal K}$.
We have eliminated the $U(1)$ gauge field by its equation of motion, thus
replacing the gauged sigma model by a sigma model with a degenerate `metric'. 
Since we postpone imposing the D-gauge, this Lagrangian contains 
the non-constant, but dependent field $e^{-{\cal K}}$
	\[
		e^{-{\cal K}} = -N_{IJ}X^I \bar{X}{}^J \;.
	\]
We perform the reduction of the Lagrangian over a time-like and space-like 
dimension simultaneously, differentiating between the two cases by
	\[
		\epsilon = 
			\begin{cases}
				-1, \;\; \text{spacelike} \;, \\
				+1, \;\; \text{timelike} \;.
			\end{cases} 	
	\]
	In order to reduce directly into the Einstein frame we decompose the metric as
	\[	
		ds^2_4 = -\epsilon\, e^{\phi} \left( dy + V_\mu dx^\mu \right)^2 + e^{-\phi} g_{\mu\nu} dx^\mu dx^\nu \;,
	\]	
	which implicitly defines $(e^{\phi},V_\mu, g_{\mu \nu})$ in terms of the four-dimensional metric $\hat{g}_{\hat{\mu} \hat{\nu}}$.
	The reduced Lagrangian is given by
	\begin{align*}
		{\cal L}_3 \sim &-\tfrac{1}{2} e^{-{\cal K}} \left( R_3 + \tfrac{1}{2} \partial_\mu \phi\, \partial^\mu \phi  - \tfrac{1}{4} \epsilon\, e^{2\phi} V^{\mu \nu} V_{\mu \nu} - \partial_\mu {\cal K} \partial^\mu \phi \right) \\
		& - e^{-{\cal K}} g_{I {J}} \partial_{\mu} X^I \partial^{\mu} \bar{X}{}^J + \tfrac{1}{4} e^{-{\cal K}} \partial_{{\mu}} {\cal K} \partial^{{\mu}} {\cal K}\\
		&+ \tfrac{1}{4}e^{\phi} {\cal I}_{IJ} ( F^I_{\mu\nu} - 2 \partial_{[ \mu} \zeta^I V_{\nu ]} ) ( F^{J \mu \nu} - 2\partial^{[ \mu} \zeta^J V^{\nu ]} ) \\ 
		& - \tfrac{1}{2}\epsilon\, e^{-\phi} {\cal I}_{IJ} \partial_\mu \zeta^I \partial^\mu \zeta^J - \tfrac{1}{2} \epsilon\, {\cal R}_{IJ} \varepsilon^{\mu \nu \rho} F^I_{\mu \nu} \partial_\rho \zeta^J \;,
	\end{align*}
	where the terms descending from the four-dimensional 
metric appear in the first line, the four-dimensional scalars in the second line, and the gauge fields in the third and fourth line.	We have denoted the field strength of the Kaluza Klein-vector by $V_{\mu \nu}$, and the scalar fields $\zeta^I = A^I_0 \, (A^I_3)$ are the components of the four dimensional vectors along the reduced timelike (spacelike) direction. The Lagrangian at present still contains the bare Kaluza Klein-vector $V_\mu$, which prevents the associated abelian  gauge symmetry from being manifest. 
Therefore we make the field redefinition
	\[
		(A^I_\mu)' := A^I_\mu - \zeta^I V_\mu \;, \hspace{2em} \Longrightarrow \hspace{2em} (F^I_{\mu \nu})' + \zeta^I V_{\mu \nu} = F^I_{\mu \nu} - 2\partial_{[ \mu} \zeta^I V_{\nu ]} \;.
	\]
	The Lagrangian now takes the manifestly gauge invariant form, 
	\begin{align*}
		{\cal L}_3 \sim &-\tfrac{1}{2} e^{-{\cal K}} \left( R_3  - \tfrac{1}{4} \epsilon\, e^{2\phi} V^{\mu \nu} V_{\mu \nu} \right)  \\
		&- e^{-{\cal K}} g_{I {J}} \partial_{\mu} X^I \partial^{\mu} \bar{X}{}^J - \tfrac{1}{4} e^{-{\cal K}}(\partial_\mu \phi - \partial_\mu {\cal K}) (\partial^\mu \phi - \partial^\mu {\cal K}) +  \tfrac{1}{4} e^{-{\cal K}} \partial_\mu {\cal K} \partial^\mu {\cal K}\\
		&+ \tfrac{1}{4}e^{\phi} {\cal I}_{IJ} ( F^I_{\mu\nu} + \zeta^I V_{\mu \nu} ) ( F^{J \mu \nu}  + \zeta^J V^{\mu \nu}) \\ 
		&- \tfrac{1}{2}\epsilon\, e^{-\phi} {\cal I}_{IJ} \partial_\mu \zeta^I \partial^\mu \zeta^J - \tfrac{1}{2} \epsilon\, {\cal R}_{IJ} \varepsilon^{\mu \nu \rho} (F^I_{\mu \nu} + \zeta^I V_{\mu \nu}) \partial_\rho \zeta^J \;,
	\end{align*}	
	where we have dropped the primes and gathered together like terms.

\subsubsection*{Conformal rescaling}
	
	In order to obtain a canonical Einstein-Hilbert term we perform the conformal rescaling
	\[
		g_{\mu \nu} =  e^{2{\cal K}}\tilde{g}_{\mu\nu} \;.
	\] 
	The various terms in the Lagrangian have the following transformation rules in three dimensions: 
	\begin{align*}
		\sqrt{g} &= \sqrt{\tilde{g}} e^{3{\cal K}} \\
		\sqrt{g}\,g^{\mu\nu} &= \sqrt{\tilde{g}}\,\tilde{g}^{\mu\nu} e^{{\cal K}} \\
		\sqrt{g}\,g^{\mu\nu}g^{\rho\sigma} &= \sqrt{\tilde{g}}\,\tilde{g}^{\mu\nu}\tilde{g}^{\rho\sigma} e^{-{\cal K}} \\
		R_3 &= e^{-2{\cal K}}\left[ \tilde{R}_3 - 4\tilde{g}^{\mu \nu}\tilde{\nabla}_\mu \tilde{\nabla}_\nu {\cal K} + 2\tilde{g}^{\mu\nu} \partial_\mu {\cal K} \partial_\nu {\cal K} \right] \;.
	\end{align*}
	The Lagrangian itself becomes
	\begin{align*}
		{\cal L}_3 \sim &-\tfrac{1}{2} \tilde{R}_3  + \tfrac{1}{8} \epsilon\, e^{2(\phi - {\cal K})} V^{\mu \nu} V_{\mu \nu}  \\
		&- g_{I{J}} \partial_{\mu} X^I \partial^{\mu} \bar{X}{}^J - \tfrac{1}{4} (\partial_\mu \phi - \partial_\mu {\cal K}) (\partial^\mu \phi - \partial^\mu {\cal K}) -  \tfrac{1}{2}  \partial_\mu {\cal K} \partial^\mu {\cal K}\\
		&+ \tfrac{1}{4}e^{(\phi - {\cal K})} {\cal I}_{IJ} ( F^I_{\mu\nu} + \zeta^I V_{\mu \nu} ) ( F^{J \mu \nu}  + \zeta^J V^{\mu \nu}) \\ 
		&- \tfrac{1}{2}\epsilon\, e^{({\cal K} - \phi)} {\cal I}_{IJ} \partial_\mu \zeta^I \partial^\mu \zeta^J - \tfrac{1}{2} \epsilon\, {\cal R}_{IJ} \varepsilon^{\mu \nu \rho} (F^I_{\mu \nu} + \zeta^I V_{\mu \nu}) \partial_\rho \zeta^J \;.
	\end{align*}	
	One can see that by redefining the KK-scalar $\phi' = \phi - {\cal K}$, the field ${\cal K}$ decouples from all other fields besides gravity. We will now set this field to be constant, and drop the primes.\footnote{When computing the tensor $g_{IJ}$, it is understood that ${\cal K}$ is
set constant after computing the derivatives.} 
This amounts to imposing the D-gauge. We could of course have done this
at an earlier stage, but we found it instructive to demonstrate how the 
radial degree of freedom ${\cal K}$ of the cone $N$ decouples.

\subsubsection*{Dualization of vector fields}
	
	Since we are working in three dimensions, and the vector fields in the Lagrangian only appear via their field strengths, it is possible to dualise the vector fields into scalar fields $(A^I , V) \sim (\tilde{\zeta}_I, \tilde{\phi})$. This is achieved by adding the Lagrange multiplier 
	\[
		{\cal L}_{\text{Lm}} \sim \tfrac{1}{2}\epsilon\, \varepsilon^{\mu \nu \rho} (F^I_{\mu \nu} \partial_\rho \tilde{\zeta}_I - V_{\mu \nu} \partial_\rho (\tilde{\phi} - \tfrac{1}{2}\zeta^I \tilde{\zeta}_I)) \;.
	\]
	The variation of ${\cal L}_3\, + \, {\cal L}_{\text{Lm}}$ gives the algebraic equations of motion (note that $\varepsilon^{\mu \nu \rho}\varepsilon_{\mu \nu \sigma} = 2\epsilon\, \delta^\rho_\sigma$)
	\begin{align*}
		V_{\mu \nu} &= 2e^{-2\phi} \varepsilon_{\mu\nu\rho}(\partial^\rho \tilde{\phi} +\tfrac{1}{2}(\zeta^I \partial^\rho \tilde{\zeta}_I - \tilde{\zeta}_I \partial^\rho \zeta^I)) \;,\\  
		F^I_{\mu \nu} &= -\epsilon\,e^{-\phi} {\cal I}^{IJ}\varepsilon_{\mu \nu \rho}(\partial^\rho \tilde{\zeta}_J - {\cal R}_{JK} \partial^\rho \zeta^K) - \zeta^I V_{\mu \nu}\;.
	\end{align*}
	Substituting the above expressions back into $\tilde{{\cal L}}_3 = {\cal L}_3\, + \, {\cal L}_{\text{Lm}}$ we are left with the dual Lagrangian
	\begin{eqnarray} \label{LagX}
		\tilde{{\cal L}}_{3} &\sim &-\tfrac{1}{2} \tilde{R}_3   - g_{I \bar{J}} \partial_{\mu} X^I \partial^{\mu} \bar{X}{}^J - \tfrac{1}{4} \partial_\mu \phi \partial^\mu \phi \\
	&&- e^{-2\phi} \left( \partial_\mu \tilde{\phi} + \tfrac{1}{2}(\zeta^I \partial_\mu \tilde{\zeta}_I - \tilde{\zeta}_I \partial_\mu \zeta^I) \right)^2 \nonumber \\ 
	&&- \tfrac{1}{2} \epsilon\, e^{-\phi}  \Big[ {\cal I}_{IJ} \partial_\mu \zeta^I \partial^\mu \zeta^J + {\cal I}^{IJ}\left( \partial_\mu \tilde{\zeta}_I - {\cal R}_{IK} \partial_\mu \zeta^K \right)\left( \partial^\mu \tilde{\zeta}_J - {\cal R}_{JL} \partial^\mu \zeta^L \right) \Big] \;. \nonumber 
	\end{eqnarray}

\subsection{A field redefinition}

We would now like to bring the Lagrangian into a form that
resembles (\ref{3dRigidLagrangian}) more closely. From \cite{Cortes:2011aj}
we know that by setting 
$(\hat{q}^a)=\frac{1}{2}(\zeta^I, 
\tilde{\zeta}_I)$ and using the real tensor $\hat{H}_{ab}$, the terms 
in the third line of (\ref{LagX}) are proportional to 
$\frac{1}{H} \hat{H}_{ab} \partial_\mu \hat{q}^a \partial^\mu \hat{q}^b$. 
\footnote{
Actually, in \cite{Cortes:2011aj} the dual coordinates $\hat{q}_a$ and 
the inverse metric $\hat{H}^{ab}$ were used, but this is simply a different
parametrization.} 
Using (\ref{eq:Hess_M_identity}) we can express this in terms of 
the Hessian metric
$\tilde{H}_{ab}$ up to  model independent terms. 
To proceed, we need to re-organize the remaining scalars 
$X^I, \phi, \tilde{\phi}$ into $2n+2$ real scalars $q^a$ which
transform as a symplectic vector and balance the $2n+2$ real scalars
$\hat{q}^a$. The counting of degree of freedom works out, because the
$n+1$ complex scalars are subject to two conditions, and therefore
correspond to $2n$ independent real scalar fields. Moreover, by going
to special real coordinates on $N$, we can relate them to a symplectic
vector. But what about $\phi$ and $\tilde{\phi}$? 

We proceed by making use of an observation that was made in 
the context of the local $r$-map, which relates the scalar 
manifolds of five-dimensional and four-dimensional vector 
multiplets \cite{Mohaupt:2009iq}.  There it is possible to 
absorb the Kaluza Klein-scalar into the manifold parametrized by the higher-dimensional
(in the case of the $r$-map, the five-dimensional) scalars. 
This amounts to lifting the constraint
imposed by the D-gauge. The Kaluza Klein-scalar is identified with the 
radial direction of the cone $N$ over $S$,
which is promoted from a gauge degree of freedom to a dynamical
degree of freedom. This idea can
be implemented in the four-dimensional setting by defining a new
set of complex scalars $Y^I$ by   
	\[
		Y^I = e^{\frac{\phi}{2}} X^I \;, \hspace{3em} \bar{Y}{}^I = e^{\frac{\phi}{2}} \bar{X}{}^I \;.
	\]
	The Kaluza Klein-scalar is now a dependent field, determined by the expression 
	\begin{equation}
	e^\phi = -i(Y^I \bar{F}_I - F_I \bar{Y}{}^I) \;.
	\end{equation}
	Since $\phi$ transforms by a shift under the global isometry group, we find that the new scalar fields must transform by a scale factor under these isometries
	\[
		Y^I \longrightarrow e^{\frac{\lambda}{2}} Y^I \;, \hspace{3em} \bar{Y}{}^I \longrightarrow e^{\frac{\lambda}{2}} \bar{Y}{}^I \;. 
	\]
	The Lagrangian can now be written as
	\begin{eqnarray}\label{LagY}
		\tilde{{\cal L}}_{3} &\sim &-\tfrac{1}{2} \tilde{R}_3   
- g_{I{J}} \partial_{\mu} Y^I \partial^{\mu} \bar{Y}{}^J - \tfrac{1}{4} \partial_\mu \phi \, \partial^\mu \phi \\
	&&- e^{-2\phi} \left[ \partial_\mu \tilde{\phi} + \tfrac{1}{2}(\phi^I \partial_\mu b_I - b_I \partial_\mu \phi^I) \right]^2 \nonumber \\ 
	&&- \tfrac{1}{2} \epsilon\, e^{-\phi}  \Big[ {\cal I}_{IJ} \partial_\mu \phi^I \partial^\mu \phi^J + {\cal I}^{IJ}\left( \partial_\mu b_I - {\cal R}_{IK} \partial_\mu \phi^K \right)\left( \partial^\mu b_J - {\cal R}_{JL} \partial^\mu \phi^L \right) \Big] \;,\nonumber
	\end{eqnarray}	
	where $\phi$ is a dependent field. 
	
	The Lagrangian is still invariant under local $U(1)$ 
transformations of the fields $(Y,\bar{Y})$, and the equations of motion transform by an overall phase factor. This is shown using that the
tensor $g_{IJ}$ has two null directions
	\[
		Y^I g_{I\bar{J}} = 0 = g_{I\bar{J}}\bar{Y}{}^J \;.	
	\]
By differentiation we obtain the identities
	\begin{align*}
		Y^I \partial_K g_{I\bar{J}} &= -g_{K\bar{J}} \;, &  \partial_K g_{I\bar{J}} \bar{Y}{}^J &= 0 \;, \\ 
		y^I \partial_{\bar{K}} g_{I\bar{J}} &= 0 \;, & \partial_{\bar{K}} g_{I\bar{J}} \bar{Y}{}^J &= - g_{I\bar{K}} \;.
	\end{align*}
	Under phase transformations the derivatives of the metric transform by a phase and the Kaluza Klein-scalar is invariant
	\begin{align*}
		\partial_K g_{I\bar{J}} &\longrightarrow e^{-i\alpha} \partial_K g_{I\bar{J}} \;, & \phi &\longrightarrow \phi \;, \\
		\partial_{\bar{K}} g_{I\bar{J}} &\longrightarrow e^{i\alpha} \partial_{\bar{K}} g_{I\bar{J}} \;. & &
	\end{align*}
	It follows that the Lagrangian is $U(1)$ invariant, the $(Y,\bar{Y})$ equations of motion transform by an overall phase, and all other equations of motion are invariant.

When comparing (\ref{LagX}) to (\ref{LagY}) both Lagrangians look identical,
except that $X^I$ have been replaced by $Y^I$. It is instructive to check
that substituting $X^I = e^{-\phi/2} Y^I$ into (\ref{LagX}) gives
indeed (\ref{LagY}). Due to the peculiar properties of the degenerate
tensor $g_{IJ}$, no derivative terms involving $\phi$ are 
generated by the substitution. All factors $e^{\phi/2}$ cancel,
because $g_{IJ}$ is homogeneous of degree $-2$, and because 
${\cal I}_{IJ}$ and ${\cal R}_{IJ}$ are homogeneous of degree 0. 
Of course the essential difference between (\ref{LagX}) and (\ref{LagY}) is
that $\phi$ is now a dependent field.

One might wonder whether the dualized Kaluza Klein-vector $\tilde{\phi}$ could
be treated in a similar way as the Kaluza Klein scalar $\phi$. 
This is not so, because the symmetries
carried by the reduced gravitational degrees of freedom $\phi,\tilde{\phi}$
do not match with the symmetries of the affine special K\"ahler manifold
$N$. The fields $\phi,\tilde{\phi}$ parametrize the coset
\[
M_{\phi, \tilde{\phi}} = \frac{SL(2,\mathbbm{R})}{SO(2)} 
\]
with isometry group $SL(2,\mathbbm{R})$. A two-dimensional solvable
subgroup generated by shifts in $\phi$ and $\tilde{\phi}$ extends
to a symmetry of the full Lagrangian once the $\hat{q}^a$ are included.
In contrast, the manifold $N$ has a homothetic action of $\mathbbm{C}^*$.
Upon taking the logarithm of the Hesse potential, the dilatation 
becomes an isometry (rather than a homothety) of the `metric' 
$g_{IJ}$.\footnote{This works as in \cite{Mohaupt:2009iq}: if the
Hesse potential is homogeneous, and we take its logarithm
as the new Hesse potential, then 
the new metric is homogeneous of degree zero (as 
a tensor, i.e. the metric coefficients are homogeneous of degree -2)
irrespective of the degree of homogeneity of the original Hesse
potential.} 
Above we observed that the fields $Y^I$ transform under shifts 
of $\phi$, and we might think of these continuous global symmetries
as residual symmetries left after we have eliminated the local dilatation 
symmetry by absorbing the KK scalar into $N$. The fields $Y^I$ are 
still subject to $U(1)$ gauge transformations, and one is tempted to
identify $\tilde{\phi}$ with the $U(1)$ gauge degree of
freedom. If this was the case one could absorb $\tilde{\phi}$
into the $Y^I$, thus making the gauge degree of freedom a physical
one. However, the global
continuous shift symmetry of $\tilde{\phi}$ do act differently 
from $U(1)$ gauge transformations, and therefore there is no
way of absorbing $\tilde{\phi}$ into $Y^I$ such that the new
variable transforms naturally under the global symmetry.
Therefore we 
proceed differently, by keeping $\tilde{\phi}$ as an independent 
field, and, consequently, keeping the local $U(1)$ gauge invariance.
We will see later that when we construct solutions, the local $U(1)$ 
gauge symmetry is gauge fixed while preserving symplectic covariance
and the isometries of scalar metric. We will also see that for 
our solutions it will always be 
possible to express $\tilde{\phi}$ in terms of the other fields. 

We can interpret our treatment of the scalar fields geometrically 
as follows.
If we freeze the scalars $\hat{q}^a$ descending from the four-dimensional
gauge fields, then the scalar manifold parametrized by the
physical four-dimensional scalars $z^i$, and by $\phi$ and $\tilde{\phi}$
is
\[
M_{z,\phi,\tilde{\phi}} = \bar{N} \times \frac{SL(2,\mathbbm{R})}{SO(2)} \;.
\]
Using the gauge equivalence, we can describe $\bar{N}$ in terms
of the fields $X^I$ using the K\"ahler quotient:
\[
M_{z,\phi,\tilde{\phi}} =
N//U(1) \times \frac{SL(2,\mathbbm{R})}{SO(2)} \;.
\]
Now we absorb $\phi$ into $N$. This `un-does' the D-gauge,
and re-institutes the radial degree of freedom of the cone
$N$ over $S$, while leaving the $U(1)$ isometry intact. The
coset $SL(2\mathbbm{R})/SO(2)$ is broken up and the remaining
one-dimensional piece is 
parametrized by the scalar $\tilde{\phi}$, with a metric
depending on $\phi$. The scalar manifold can be 
represented as a deformed product
\[
M_{z,\phi,\tilde{\phi}} =
N/U(1) \times_\phi \mathbbm{R}_{\tilde{\phi}} \;.
\]
Here $N/U(1)$ is the quotient of $N$ with respect to its
$U(1)$ isometry rather than the K\"ahler quotient. 
The advantage of this way of organising the fields becomes
apparent once we use real special coordinates on $N$.

\subsection{The real parametrization}
	
	The kinetic term of the complex scalar 
fields takes precisely the same form as considered previously in section \ref{Sect:Real}  
	\[
		{\cal L}_{3} \sim - g_{I{J}} \partial_{\mu} Y^I \partial^{\mu} \bar{Y}{}^J + \cdots\;.
	\]
	We make the real decomposition
	\[
		Y^I = x^I + i u^I(x,y) \hspace{3em} F_I = y_I + iv_I(x,y) \;, 
	\]
	and use the previous results to write this term in the Lagrangian as
	\begin{equation}
\label{L3_real_scalar}
		{\cal L}_{3} \sim - \left[ \tilde{H}_{ab} - \frac{1}{4H^2}H_a H_b + \frac{1}{H^2}(\Omega_{ac}q^c)(\Omega_{bd}q^d) \right] \partial_\mu q^a \partial^\mu q^b + \cdots \;,
	\end{equation}
	where $q^a = (x^I, y_I)^T$. Note that our previous calculations
are still applicable after the replacement $X^I \rightarrow
Y^I$ , due to homogeneity.

The Kaluza Klein-scalar is given in terms of the real variables by
	\begin{equation}
\label{eq:phi_H}
		e^\phi = -2H = -2(x^I v_I(x,y) - y_Iu^I(x,y)) \;,
	\end{equation}
	which is homogeneous of degree two in $q^a = (x^I, y_I)^T$. The kinetic term for the Kaluza Klein-scalar can then be written as
	\[
		\tfrac{1}{4}\partial_\mu \phi \, \partial^\mu \phi = \frac{1}{4H^2}H_aH_b \partial_\mu q^a \partial^\mu q^b \;,
	\]
and this term cancels against the second term in (\ref{L3_real_scalar}).
When rewriting the terms descending from the four-dimensional gauge fields using the
variables $\hat{q}^a = (\frac{1}{2}\zeta^I, \frac{1}{2}\tilde{\zeta}_I)^T$, 
they take the
form
	\[
		{\cal L}_{3\text{gauge}} \sim \epsilon\, \tilde{H}_{ab} \partial_\mu \hat{q}^a \partial^\mu \hat{q}^b + \epsilon\, \frac{2}{H^2} \left( q^a \Omega_{ab} \partial_\mu \hat{q}^b \right)^2  - \frac{1}{4H^2} \left[ \partial_\mu \tilde{\phi} - 2\left(\hat{q}^a \Omega_{ab} \partial_\mu \hat{q}^b \right) \right]^2 \;. 
	\]
	We can now put together all terms and write the Lagrangian in terms of real fields as 
	\begin{eqnarray}
			\tilde{{\cal L}}_3 &\sim  &- \tfrac{1}{2} \tilde{R}_3 - \tilde{H}_{ab} \left(\partial_\mu q^a \partial^\mu q^b - \epsilon\, \partial_\mu \hat{q}^a \partial^\mu \hat{q}^b \right)  \nonumber \\ 
			&&- \frac{1}{H^2} \left( q^a \Omega_{ab} \partial_\mu q^b \right)^2 + \epsilon\, \frac{2}{H^2} \left( q^a \Omega_{ab} \partial_\mu \hat{q}^b \right)^2  \nonumber \\
			&&- \frac{1}{4 H^2} \left( \partial_\mu \tilde{\phi} + 2\hat{q}^a \Omega_{ab} \partial_\mu \hat{q}^b  \right)^2 \;. \label{eq:Lag_Hess} 
 	\end{eqnarray}
This formula is one of our main results, and provides a 
new formulation of the supergravity $c$-map (and its temporal
version) in terms of real variables and the Hesse potential.
It comes surprisingly close to the Sasaki-type form of the
rigid and local $r$-map and rigid $c$-map. The scalar term
in the first line has precisely the form found for the local
$r$-map, a Sasaki-type metric with the Hesse potential of the
rigid theory being replaced by its logarithm. The terms in 
second and third line are simple and universal, they only
depend on the constant matrix $\Omega_{ab}$ and the Hesse 
potential $H$ (identified with the Kaluza-Klein scalar). We 
can also understand the origin of these additional terms.
First, there is one term involving the dualized Kaluza-Klein
vector $\tilde{\phi}$. This field plays a special role because
we could not absorb it into $N$ in the same way as the
Kaluza-Klein scalar. The other terms can be understood from
our real formulation of projective special K\"ahler geometry.
They arise from rewriting the tensor fields $H^{(0)}_{ab}$ and
$\hat{H}_{ab}$ in terms of $\tilde{H}_{ab}$. In the analogous
case of the $r$-map such terms are absent, because there
the analogues of $\tilde{H}_{ab}$ and $\hat{H}_{ab}$ coincide,
and because the scalar metric becomes the analogue of $\tilde{H}_{ab}$
after absorbing the Kaluza-Klein scalar.

The fields in (\ref{eq:Lag_Hess}) 
are still subject to $U(1)$ gauge transformations,
and therefore the quaternion-K\"ahler metric on the physical
scalar manifold is obtained by a $U(1)$ quotient. One could 
impose a gauge fixing condition and eliminate one of the scalar
fields. 
Since the metric on the $U(1)$ bundle parametrized
by $q^a, \hat{q}^a, \tilde{\phi}$ is degenerate along the direction
generated by the $U(1)$, we can choose any condition which is
transverse to the $U(1)$ action (such as $q^1=0$)  
and then restrict the (degenerate) metric 
on the bundle to
the resulting hypersurface to obtain the positive definite 
quaternion-K\"ahler metric (or split signature para-Quaternion-K\"ahler
metric). Since the $U(1)$ action relates the members of the $S^1$ family
of special connections to one another, the $U(1)$ bundle can be viewed
as the bundle of special connections, and a $U(1)$ gauge fixing as
picking a special connection.

We prefer not to fix the $U(1)$ gauge and to work on the $U(1)$ bundle,
because, as we explained in section \ref{Sec:LocalVM}, a $U(1)$ gauge fixing
would spoil the manifest symplectic covariance. In the following
section we will show that instantonic solutions can be constructed 
and be lifted to solitonic solutions, such as black holes, while 
preserving symplectic covariance. We will then revisit the issue
of $U(1)$ gauge fixing.



\section{Stationary Solutions}

	We now turn to finding stationary solutions of the four-dimensional Lagrangian. Four-dimensional stationary BPS solutions for general
vector multiplet couplings 
have been constructed some time ago by imposing invariance under 
part of the supersymmetry transformations 
\cite{Bergshoeff:1996gg,Behrndt:1997ny,Denef:2000nb,LopesCardoso:2000qm}. We expect to
recover these solutions and to obtain further non-BPS solutions.
To this end we reduce over a time-like dimension, therefore making the choice $\epsilon = 1$ in the formula for the reduced Lagrangian 
(\ref{eq:Lag_Hess}). We will find that in flat backgrounds we can give 
solutions to generic models in terms of harmonic functions.

Before embarking into the details, let us explain the overall strategy. 
Since the three-dimensional Lagrangian is a combination of perfect
squares, we will try to reduce the field equations to Bogomol'nyi
equations which follow from imposing that the squares vanish individually. 
We will focus on solutions where the three-dimensional metric is
Ricci-flat, and, hence, flat. This restricts the fields to take 
values in totally isotropic submanifold, and therefore we will call
the corresponding ansatz the isotropic ansatz.
After lifting to four dimensions we will 
obtain four-dimensional
extremal static black hole solutions as well as over-extremal (singular)
rotating solutions. 
The structure of the
Bogomol'nyi equations can be read off from the Lagrangian
(\ref{eq:Lag_Hess}). One of the Bogomol'nyi equations results
from imposing that the first line of (\ref{eq:Lag_Hess}) vanishes,
which gives a relation of the form
\[
\partial_\mu q^a = \pm \partial_\mu \hat{q}^a
\]
between the $q^a$ and the $\hat{q}^a$, which 
is identical to the relation found for five-dimensional black 
holes \cite{Mohaupt:2009iq}. 
If the scalar metric satisfies a
certain compatibility condition, 
one can instead impose the more
general condition
\[
\partial_\mu q^a = R^a_{\;\;b} \partial_\mu \hat{q}^b \;,
\]
where $R^a_{\;\;b}$ is a constant `field rotation matrix'. Such solutions
are non-BPS, and will be discussed in a separate section.
Once either of these condition is imposed, 
the terms in the second line combine into one term,
which, however, has a similar structure as a term within the
square in the third line.
The most  general ansatz only requires that
the second and third line vanish in combination, 
while a more restricted ansatz requires
that the second and third line vanish independently. The restricted 
ansatz corresponds to static solutions, because imposing that the
third line vanishes is equivalent to the vanishing of the field
strength of the Kaluza-Klein vector. Without this restriction,
we obtain stationary rotating solutions.
We will refer to solutions obtained from our isotropic ansatz
as isotropic solutions. Note that they will in general neither
be BPS (since we admit a field rotation matrix), nor extremal
(since rotating solutions are over-extremal).

As in the five-dimensional case \cite{Mohaupt:2009iq}, 
we will be able to demonstrate that
the equations of motion can be reduced to decoupled harmonic equations
by choosing suitable `dual' coordinates. Therefore the solution will be 
given in terms of a set of harmonic functions. We will also see that
this way we naturally obtain the generalized stabilization equations
of four-dimensional black holes, in their algebraic and manifestly
symplectic form.

\subsection{Equations of motion}

We will now derive all the field equations of the Lagrangian
(\ref{eq:Lag_Hess}) and show explicitly how they are solved by
imposing Bogomol'nyi equations.

	Firstly, we perform the variation of the equation (\ref{eq:Lag_Hess}) with respect to the field $q^a$ to obtain the equation of motion
	\begin{align}
			&2\nabla^\mu \left[\tilde{H}_{ab}\partial_\mu q^b \right] -  \partial_a \tilde{H}_{bc} \left(\partial_\mu q^b \partial^\mu q^c - \partial_\mu \hat{q}^b \partial^\mu \hat{q}^c \right) \notag\\ 
			&+ 2\nabla^\mu \left[ \frac{1}{H^2} q^c \Omega_{ca} \left( q^d \Omega_{de} \partial_\mu q^e \right) \right]  +  \notag \\
			&- 2\partial_a \left( \frac{1}{H} q^c \right) \left[ \Omega_{cb} \partial_\mu q^b \frac{1}{H} \left( q^d \Omega_{de} \partial_\mu q^e \right) - 2 \Omega_{cb} \partial_\mu \hat{q}^b \frac{1}{H} \left( q^d \Omega_{de} \partial_\mu \hat{q}^e \right) \right] \notag \\
			&- \partial_a \left(\frac{1}{4H^2}\right)\left( \partial_\mu \tilde{\phi} + 2\hat{q}^a \Omega_{ab} \partial_\mu \hat{q}^b  \right)^2 = 0\;. \label{eq:eom1} 
	\end{align}
	Next, the equation of motion for the $\hat{q}^a$ fields
	\begin{align}
			&- 2 \nabla^\mu \left[\tilde{H}_{ab}\partial_\mu \hat{q}^b \right] \notag  \\ 
			&- 4 \nabla^\mu \left[ \frac{1}{H^2} q^c \Omega_{ca} \left( q^d \Omega_{de} \partial_\mu \hat{q}^e \right) \right] + \nabla^\mu \left[ \frac{1}{H^2} \hat{q}^b \Omega_{ba} \left( \partial_\mu \tilde{\phi} + 2\hat{q}^c \Omega_{cd} \partial_\mu \hat{q}^d  \right) \right]  \notag \\
			&- \frac{1}{H^2}\Omega_{ab} \partial_\mu \hat{q}^b \left( \partial_\mu \tilde{\phi} + 2\hat{q}^c \Omega_{cd} \partial_\mu \hat{q}^d  \right) = 0\;. \label{eq:eom2}
	\end{align}
	The equation of the field $\tilde{\phi}$, which descends from the Kaluza Klein-vector, is given by
	\begin{equation}
		\nabla^\mu \left[ \frac{1}{4H^2}\left(\partial_\mu \tilde{\phi} + 2\hat{q}^c \Omega_{cd} \partial_\mu \hat{q}^d \right) \right] = 0 \;.
		\label{eq:eom3}
	\end{equation}
	This equation is nothing but the Bianchi identity for $V_{\mu\nu}$, the field strength of the Kaluza Klein-vector, which allow us to write the field strength in terms of a gauge potential $V_{\mu \nu} = \partial_\mu V_\nu - \partial_\nu V_\mu$.
	Finally, from the variation of the metric we find the Einstein equations
	\begin{align}
			&- \tfrac{1}{2} \tilde{R}_{3\mu\nu} - \tilde{H}_{ab} \left(\partial_\mu q^a \partial_\nu q^b - \partial_\mu \hat{q}^a \partial_\nu \hat{q}^b \right)  \notag \\ 
			&- \frac{1}{H^2} \left( q^a \Omega_{ab} \partial_\mu q^b \right)\left( q^c \Omega_{cd} \partial_\nu q^d \right) + \frac{2}{H^2} \left( q^a \Omega_{ab} \partial_\mu \hat{q}^b \right)\left( q^c \Omega_{cd} \partial_\nu \hat{q}^d \right)  \notag\\
			&- \frac{1}{4 H^2} \left( \partial_\mu \tilde{\phi} + 2\hat{q}^a \Omega_{ab} \partial_\mu \hat{q}^b \right)\left( \partial_\nu \tilde{\phi} + 2\hat{q}^c \Omega_{cd} \partial_\nu \hat{q}^d  \right)  = 0\;. \label{eq:eom_Einstein} 
	\end{align}

\subsubsection*{Dual coordinates}

	The Hessian matrix $\tilde{H}_{ab}$ allows us to define a natural set of dual coordinates
	\begin{align*}
		q_a &:= \tilde{H}_a = -\tilde{H}_{ab} q^b  \\
			&= -\frac{H_a}{2H} = \frac{1}{H} \left( \begin{array}{c} -v_I \\ u^I \end{array} \right) \;.
	\end{align*}
	By the chain rule we find the expression for the derivative of the dual coordinates
	\[
		\tilde{H}_{ab} \partial_\mu q^b = \partial_\mu q_a = \partial_\mu \left[ \frac{1}{H} \left( \begin{array}{c} -v_I \\ u^I \end{array} \right) \right] \;.
	\]
	The existence of these dual coordinates is critical for obtaining solutions to generic models in terms of harmonic functions. Note that the definition 
of the dual coordinates is completely analogous to the five-dimensional
case \cite{Mohaupt:2009iq}.

\subsection{The isotropic ansatz}

A flat three-dimensional geometry requires that the energy-momentum tensor must vanish identically. To achieve this we must impose an appropriate ansatz for the fields, which consists of two distinct parts. The first part of our ansatz is to identify the vectors $\partial_\mu q^a$ and $\partial_\mu \hat{q}^a$ 
up to an overall sign
	\begin{equation}
		\partial_\mu q^a = \pm \partial_\mu \hat{q}^a \;. \label{eq:ansatz1}
	\end{equation}
Upon imposing this ansatz the vacuum Einstein equations reduce to 
\[	
	\frac{1}{4H^2}\left(\partial_\mu \tilde{\phi} + 2\hat{q}^q \Omega_{ab} \partial_\mu q^b \right)^2 = \frac{1}{H^2}\left(q^a \Omega_{ab} \partial_\mu q^b\right)^2 \;.
\]
The second part of our ansatz is now clear: we must make the identification
\begin{equation}
	\tfrac{1}{2} \left( \partial_\mu \tilde{\phi} + 2\hat{q}^a \Omega_{ab} \partial_\mu \hat{q}^b \right) = q^a\Omega_{ab}\partial_\mu q^b \;,
	\label{eq:ansatz2}
\end{equation}
	where the choice of sign is important.
	One can interpret this as fixing $\tilde{\phi}$ in terms of other fields which are independent
	\[
		\tilde{\phi} = 2(q^a \mp \hat{q}^a)\Omega_{ab}q^b\;.
	\]
	Note that our first ansatz means that $q^q \mp \hat{q}^a$ is a constant in spacetime.
	By construction the ans\"atze (\ref{eq:ansatz1}) and (\ref{eq:ansatz2}) solve the Einstein equations with a flat spacetime metric. This means that the scalar fields take values in a totally isotropic submanifold of the target space of the non-linear sigma model described by the Lagrangian (\ref{eq:Lag_Hess}).

	Next, we need to consider the effect this ansatz has on the other equations of motion. Firstly, from the $\tilde{\phi}$ equation of motion we find the condition
	\begin{equation}
		\nabla^\mu\left[ \frac{1}{H^2} \left(q^a\Omega_{ab}\partial_\mu q^b \right) \right] = 0 \;. \label{eq:int_phi_eom}
	\end{equation}
	Turning our attention to the $q^a$ equation of motion, we see that the second term will drop out, the third line will simplify, and due to (\ref{eq:int_phi_eom}) the derivative in the second line will only act on $q^c$. We are left with
\begin{align*}
	&2\nabla^\mu \left[ \tilde{H}_{ab} \partial_\mu q^b \right] + \\
	&+ \frac{2}{H^2}\partial_\mu q^c \Omega_{ca} \left(q^d\Omega_{de}\partial_\mu q^e \right) + 2 \partial_a \left(\frac{1}{H}q^c \right)\Omega_{cb}\partial_\mu q^b\frac{1}{H}\left( q^d\Omega_{de}\partial_\mu q^e\right) \\
	&- 2\partial_a \left(\frac{1}{H} \right) q^c\Omega_{cb}\partial_\mu q^b \frac{1}{H}\left( q^d\Omega_{de}\partial_\mu q^e\right) = 0\;.
\end{align*}
 The fourth term then cancels with the derivative acting on the Hesse potential in the third term 
\[
	2\nabla^\mu \left[ \tilde{H}_{ab} \partial_\mu q^b \right]+ \left(\frac{2}{H^2}\partial_\mu q^c \Omega_{ca} + 2 \frac{1}{H^2}\Omega_{ab}\partial_\mu q^b \right) \left(q^d\Omega_{de}\partial_\mu q^e \right)	 = 0\;.
\]
Since $\Omega_{ab}$ is antisymmetric the second term cancels with the third term, and writing the first term in terms of the dual coordinates $q_a$  we are finally left with
\begin{equation}
	\Delta q_a = 0 \;. \label{eq:harm_func}
\end{equation}
This is the Laplace equation for the dual coordinates $q_a$, with respect to the flat Euclidean three-dimensional metric. Solutions are given by harmonic functions.
	
	Now let us consider the $\hat{q}^a$ equation of motion. From (\ref{eq:int_phi_eom}) we see that the derivative in the second term will only act on $q^c$, and the second and third term will simplify to give
	\begin{align*}
		&-2\nabla^\mu \left[ \tilde{H}_{ab} \partial_\mu \hat{q}^b \right] -\left( \frac{2}{H^2}\partial_\mu q^c \Omega_{ca} + \frac{2}{H^2}\Omega_{ab}\partial_\mu q^b \right) \left( q^d\Omega_{de}\partial_\mu q^e\right)  = 0\;.
	\end{align*} 
	The second and third term cancel due to antisymmetry of $\Omega_{ab}$, and we again get the Laplace equation on the dual coordinates (\ref{eq:harm_func}).
	
	Let us finally check that the solutions to the $q$ and $\hat{q}$ equations of motions are consistent with the $\tilde{\phi}$ equation of motion (\ref{eq:int_phi_eom}). 
Using the identity $q^a \Omega_{ac} = -\tfrac{1}{4}H_a \Omega^{ab} H_{bc}$,
we can write the LHS of (\ref{eq:int_phi_eom}) in terms of dual coordinates as
	\begin{align*}
		\nabla^\mu\left[ \frac{1}{H^2} \left(q^a\Omega_{ab}\partial_\mu q^b \right) \right] &= -\tfrac{1}{4}\nabla^\mu\left[ \frac{1}{H^2} \left(H_{a}\Omega^{ab} H_{bc} \partial_\mu q^c \right) \right] \\
		&= -\nabla^\mu\left[ \tilde{H}_{a}\Omega^{ab} \tilde{H}_{bc} \partial_\mu q^c  \right] \\ 		
		&= -\nabla^\mu\left[ q_{a}\Omega^{ab} \partial_\mu q_b   \right] \\
		&= -q_{a}\Omega^{ab} \Delta q_b  \;.
	\end{align*}
	It is clear that for solutions satisfying the Laplace equation the RHS will vanish. We conclude that upon imposing our ans\"atze all equations of motion reduce to the Laplace equation on the dual coordinates (\ref{eq:harm_func}).

When rewriting the isotropic ansatz (\ref{eq:ansatz1}) in terms of 
four-dimensional quantities one recovers a well known relation which 
for four-dimensional BPS solutions follows from supersymmetry.
First, it is useful to note that the three-dimensional scalars $\hat{q}^a=
\frac{1}{2}(\zeta^I,  \tilde{\zeta}_I)$ are related to four-dimensional
field strength by
\[
\partial_\mu \zeta^I = F^I_{\mu 0} \;,\;\;\;
\partial_\mu \tilde{\zeta}_I = G_{I|\mu 0} \;.
\]
While the first relation holds by definition, the second requires one to 
combine and manipulate various of the relations in this section. The
above relations show that the scalar fields $\zeta^I, \tilde{\zeta}_I$
can be interpreted as electro-static potentials for the field strength 
and Hodge-dual field strength. Combining this with  
$(q^a) = \frac{1}{2} \left(Y^I + \bar{Y}^I, F_I(Y) + \bar{F}_I(\bar{Y})\right) 
= \frac{1}{2} e^{\phi/2} \left(X^I + \bar{X}^I, F_I(X) + \bar{F}_I(\bar{X})
\right)$, the isotropic ansatz (\ref{eq:ansatz1}) becomes
\begin{align}
\partial_\mu (e^{\phi/2} (X^I + \bar{X}^I )) &= \pm F^I_{\mu 0} =
\pm (F^{I|+}_{\mu 0} + F^{I|-}_{\mu 0}) \;, \label{eq:gauge1} \\
\partial_\mu (e^{\phi/2}(F_I + \bar{F}_I )) &= \pm G_{I|\mu 0} = 
\pm (G^+_{I|\mu 0} + G^-_{I|\mu 0}) \label{eq:gauge2} \;.
\end{align}
Thus the isotropic ansatz implies that the real part of the symplectic
vector $(X^I,F_I)$ is proportional to the gauge potentials. 
For supersymmetric solutions this follows from
the gaugino variation, see for example \cite{LopesCardoso:2000qm},
while here we obtain it as the Bogomol'nyi equation associated to
the first line of (\ref{eq:Lag_Hess}).

\subsubsection*{Remarks on the local $U(1)$ symmetry}

	The ansatz (\ref{eq:ansatz1}), (\ref{eq:ansatz2}) 
for stationary BPS solutions breaks the manifest local $U(1)$ 
invariance of the equations of motion. This is obvious since we equate 
quantities which transform under the $U(1)$ to quantities which don't. 
In other words the ansatz implicitly fixes the $U(1)$ gauge.  
Since symplectic covariance and the global isometries are respected 
by the ansatz, the gauge fixing respects these symmetries. Moreover,
once the equations of motion have been solved, we can specify 
the gauge fixing condition explicitly. Re-writing the solutions 
$q_a = {\cal H}_a$, where ${\cal H}_a$ are harmonic functions, in terms of the
complex variables, this becomes
\begin{equation}
\label{Attractor1}
e^{-\phi} (Y^I - \bar{Y}^I) = -i{\cal H}^I \;,\;\;\;
e^{-\phi} (F_I - \bar{F}_I) = -i{\cal H}_I  \;,
\end{equation}
where ${\cal H}^I, {\cal H}_I$ are harmonic functions. Rewriting this
in terms of the original four-dimensional fields $X^I$, 
we obtain
\begin{equation}
\label{Attractor2}
e^{-{\phi/2}} (X^I - \bar{X}^I) = -i{\cal H}^I \;,\;\;\;
e^{-{\phi/2}} (F_I - \bar{F}_I) = -i{\cal H}_I  \;.
\end{equation}
Using the D-gauge 
$-i(X^I \bar{F}_I - F_I \bar{X}^I)=1$, we can
verify that 
\begin{equation}
\label{U1condition}
X^I {\cal H}_I - F_I {\cal H}^I = e^{-\phi/2} \;.
\end{equation}
This relation is clearly not $U(1)$ invariant and can be viewed as
the $U(1)$ gauge fixing implied by our ansatz. It reflects that the
fields $Y^I$ only correspond to $2n+1$ independent scalar fields.
This missing real scalar, the dualized Kaluza-Klein vector $\tilde{\phi}$,
is determined by its own equation of motion. Note that we could not gauge
fix the $U(1)$ by a symplectically invariant condition of the form
(\ref{U1condition}) without imposing part of the field equations.
As explained in section \ref{Sec:LocalVM} a condition of this
type forces the fields to be orthogonal to the $U(1)$ action. Since
this distribution is the contact distribution of the Sasakian, it
is not integrable, and cannot be used to realize $\bar{N}$ as a 
hypersurface in the Sasakian (or the (para-)quaternion-K\"ahler
manifold as a hypersurface in the principal bundle parametrized
by $q^a, \hat{q}^a, \tilde{\phi}$). However, solutions to the field
equations correspond to maps into lower-dimensional submanifolds
of the scalar target space, and integral manifolds of lower dimension
do exist.

We remark that there is an alternative description which allows 
to keep the $U(1)$ invariance manifest and effectively decouples
the $U(1)$ gauge degree of freedom. As in \cite{LopesCardoso:2000qm},
one can modify the definition of $Y^I$ as
\begin{equation}
Y^I = e^{\phi/2} \bar{h} X^I \;, \label{eq:mod_Y}
\end{equation}
where $h$ is a phase factor which transforms with the same charge under 
$U(1)$ as $X^I$. 
Note that in \cite{LopesCardoso:2000qm} a different convention is used, which corresponds in our notation to taking $Y^I = e^{-\phi/2} \bar{h} X^I$, instead
of the above relation. This only alters how $e^\phi$ depends on the independent coordinates $Y^I$, but has no baring on our discussion.
When comparing to \cite{LopesCardoso:2000qm}, 
note that the Kaluza-Klein scalar  $e^\phi$ is related to the functions $f,g$
used there by $e^\phi = e^{-2f} = e^{2g}$. 

The effect of the modified definition (\ref{eq:mod_Y}) is that $Y^I$ is now a $U(1)$ invariant field. 
Due to the degeneracy of $g_{IJ}$ and the homogeneity properties of
the functions involved, this modification does not change the calculations 
presented above. In particular, no derivative term for the field $h$ is
generated. When rewriting (\ref{Attractor1}), by replacing 
the $U(1)$ invariant variables $Y^I$ by the original variables
$X^I$, which are subject to $U(1)$ transformations, 
we obtain:
\[
e^{-\phi/2} (\bar{h} X^I - h \bar{X}^I) = -i {\cal H}^I \;,\;\;\;
e^{-\phi/2} (\bar{h} F_I - h \bar{F}_I) = -i {\cal H}_I \;,
\]
as in \cite{LopesCardoso:2000qm}, except for a different normalization
of the $Y^I$.\footnote{And, of course, in the present paper we do not consider
higher derivative terms.}
Using again the D-gauge $-i(X^I \bar{F}_I - F_I \bar{X}^I) = 1$, this implies
\[
X^I {\cal H}_I - F_I {\cal H}^I =  h e^{-\phi/2} \;,
\]
which determines the compensating phase $h$ for our solution.

\subsubsection*{Remarks on attractor behaviour and gradient flow equations}

The equations (\ref{Attractor1}), (\ref{Attractor2}) 
are the well known black hole
attractor equations. To be precise the term attractor equations 
is applied in the literature to both the equations which determine the values
of the scalars on the horizon, and to the more general equations
which determine the scalars globally in terms of harmonic functions.
Here we have recovered the global version, the horizon version can
be obtained by taking the near horizon limit. The equations 
(\ref{Attractor1}) are algebraic equations, and they are symplectically
covariant. Another formulation  of the attractor equations takes the form 
of gradient flow equations driven by a so-called `fake superpotential'
\cite{Ceresole:2007wx,LopesCardoso:2007ky,Perz:2008kh}.
Most of the literature on gradient flow equations focuses on spherically 
symmetric solutions and uses the physical scalars $z^i$, so that the
resulting equations are not symplectically covariant. Recently
the BPS equations for four-dimensional ${\cal N}=2$ gauge theories were
reformulated, using the Hesse potential, in symplectically covariant
form, for general non-spherical solutions \cite{VandenBleeken:2011ib}.

Our formalism by-passes the gradient flow equations and we directly
obtain solutions in terms of harmonic functions. While we leave 
a comprehensive discussion of the relation between our approach and
gradient flow equations for future work, we would like to expand
a little on the discussion given in \cite{Mohaupt:2009iq}, where we
observed that the field equation can be recast in first order form.
One way of re-writing the second order equations of motion into first order
form is to rewrite the Lagrangian as a (possibly alternating) 
sum of squares.  This can be done systematically
within our formalism, as follows. Upon inspection of the 
Lagrangian (\ref{eq:Lag_Hess}) we see that the second and third line are 
already written as the sum of square terms. We then only need to consider the 
first line, which we can write as
\begin{align*}
	\tilde{H}_{ab}(\partial_\mu q^a \partial^\mu q^b - \partial_\mu \hat{q}^a \partial^\mu \hat{q}^b) =\;\;& \tilde{H}_{ab}(\partial_\mu q^a \pm \tilde{H}^{ac} \partial_\mu {\cal H}_c) (\partial^\mu q^b \pm \tilde{H}^{bd} \partial^\mu {\cal H}_d) \\
	&- \tilde{H}_{ab}(\partial_\mu \hat{q}^a - \tilde{H}^{ac} \partial_\mu {\cal H}_c) (\partial^\mu \hat{q}^b - \tilde{H}^{bd} \partial^\mu {\cal H}_d) \\
	&+ \text{Total derivatives} \;,
\end{align*}
where ${\cal H}_a$ are harmonic functions. In the spherically symmetric case
one can dimensionally reduce the Lagrangian to one dimension, where derivatives
of harmonic functions are just constants, which can be identified with the
conserved charges carried by the solution. One then obtains gradient flow
equations, which are driven by the central charge in the supersymmetric
case and by a fake superpotential in general. We refer to \cite{Mohaupt:2009iq}
for a discussion of the spherically symmetric case and proceed without
imposing spherical symmetry. 

The first part $\partial_\mu q^a = \pm \partial_\mu \hat{q}^a$ 
of the isotropic ansatz can be seen as imposing that the squares 
displayed above vanish. The second part of the isotropic ansatz 
matches the remaining squares, which appear with a relative sign difference, 
and, hence, the sum of all squares vanishes.
The reduces the field equations of the three-dimensional
scalars to first order
equations,
which become the usual flow equations upon imposing spherical symmetry. 
By eliminating the fields $\hat{q}^a$ by their equations of motion, we
are left with (generalized) flow equations for the fields $q^a$, which 
are the four-dimensional scalars combined with the Kaluza-Klein scalar, 
i.e. a component of the four-dimensional metric. 

When we instead eliminate the harmonic functions, we recover 
the isotropic ansatz. 
We can also make contact with relations recently found 
in \cite{VandenBleeken:2011ib} by contracting 
\[
\partial_\mu q^a = \pm \partial_\mu \hat{q}^a \;. 
\]
with $q_a=\tilde{H}_a$.
Then the left-hand side is related to the gradient of the Hesse potential,
\[
\partial_\mu \tilde{H} = \tilde{H}_a \partial_\mu q^a = q_a
\partial_\mu q^a \;,
\]
while the right-hand side is
\begin{eqnarray}
q_a \partial_\mu \hat{q}^a &=& e^{-\phi/2}\left( (\mbox{Im} F_I(X)) \partial_\mu
\zeta^I - (\mbox{Im} X^I) \, \partial_\mu \tilde{\zeta}_I \right) 
\nonumber \\
&=&
e^{-\phi/2}\left( (\mbox{Im} F_I(X)) F^I_{\mu 0}  - (\mbox{Im} X^I) 
G_{I|\mu 0}  \right)\;. \nonumber 
\end{eqnarray}
This can be related to the expression for the graviphoton in terms of
real coordinates by Hodge-dualizing the field strength
\[
 q_a \partial_\mu \hat{q}^a = + \frac{1}{2} \epsilon_{0\mu \nu \rho}
e^{-\phi/2} \left( \mbox{Im} F_I \tilde{F}^{I|\nu \rho} -
\mbox{Im} X^I \tilde{G}_I^{\nu \rho} \right)  = \frac{1}{4} e^{-\phi/2}
\epsilon_{0\mu \nu \rho} T^{\nu \rho}\;.
\]
Thus we obtain a relation between the gradient of the Hesse potential
and the magnetic components of the graviphoton, or, equivalently,
the electric components of the Hodge-dual of the graviphoton
\[
\partial_\mu \tilde{H} = \pm  \frac{1}{4} e^{-\phi/2}
\epsilon_{0\mu \nu \rho} T^{\nu \rho} = \pm \frac{1}{2} e^{-\phi/2}
\tilde{T}_{0\mu} \;.
\]
This relation appears to be the local analogue of an equation for the gradient
of the Hesse potential recently found in \cite{VandenBleeken:2011ib}
for BPS dyons in rigid ${\cal N}=2$ theories. As the unique symplectically
invariant contraction between scalars and gauge fields, the 
graviphoton plays the role of the central charge vector field
used in \cite{VandenBleeken:2011ib}.

\subsection{Rotating solutions}

	We now have an ansatz for finding stationary isotropic solutions (flat 3d metric) to completely generic models in terms of the dual coordinates. However, in order to write down these solutions explicitly in terms of the four-dimensional fields one must disentangle them from the dual coordinates. This is equivalent to solving the generalised stabilisation equations, and is not always possible in closed form. In this section we will discuss solutions which lift to rotating over-extremal solutions in four-dimensions, with the STU model as an explicit example. These solutions are characterised by axial symmetry and the requirement that they are asymptotic to Minkowski space at infinity. 

	The results of the previous section show that upon imposing our isotropic ansatz (\ref{eq:ansatz1}) and (\ref{eq:ansatz2}), the equations of motion reduce to $\Delta q_a = 0$, and solutions are given in terms of the dual coordinates by harmonic functions
	\begin{equation}
		q_a = \frac{1}{H}\left( \begin{array}{c} -v_I \\ u^I \end{array} \right) = \left( \begin{array}{c} -{\cal H}_I \\ {\cal H}^I \end{array} \right)  = {\cal H}_a \;. \label{eq:solution}
	\end{equation}
	We wish to disentangle the four-dimensional metric from this solution, and show that it corresponds to a rotating solution. We can do this by retracing our steps in the dimensional reduction procedure to find 	
	\begin{equation}
		g_{\mu \nu} = \delta_{\mu \nu} \;, \hspace{2em} e^{\phi} = -2H \;, \hspace{2em} \partial_\mu V_\nu =  \tfrac{1}{2}\varepsilon_{\mu \nu \rho}\left( {\cal H}_I \partial^\rho {\cal H}^I - {\cal H}^I\partial^\rho {\cal H}_I \right) \;. \label{eq:STU_4d_metric}
	\end{equation}
	The first equation is trivial; the second is model dependent and we will look into it in more detail later. For now let us focus on the third equation, or more accurately set of equations. These are entirely independent of the details of the model, i.e.\ choice of prepotential. Following the method for producing rotating isotropic solutions used in \cite{Bergshoeff:1996gg,Behrndt:1997ny}, we impose that solutions are axially symmetry about the coordinate $\varphi$ in an oblate spheroidal coordinate system, defined by
	\begin{align*}
		x &= \sqrt{r^2 + \alpha^2} \sin \theta \cos \varphi \;, \\
		y &= \sqrt{r^2 + \alpha^2} \sin \theta \sin \varphi \;, \\
		z &= r\cos \theta \;.
	\end{align*}
	The (flat) three-dimensional Euclidean metric is given in these coordinates by
	\[
		ds^2_3 = \left( \frac{r^2 + \alpha^2 \cos^2 \theta}{r^2 + \alpha^2} \right) dr^2 + (r^2 + \alpha^2\cos^2\theta)d\theta^2 + (r^2 + \alpha^2)\sin^2\theta d\varphi^2 \;.
	\] 
	In this coordinate system the third set of equations in (\ref{eq:STU_4d_metric}) become
	\begin{align}
	\frac{1}{(r^2 + \alpha^2)\sin \theta}\partial_\theta V_\varphi &= \tfrac{1}{2} \left({\cal H}_I \partial_r {\cal H}^I - {\cal H}^I\partial_r {\cal H}_I \right) \;, \label{eq:STU_KKvector1}\\
	-\frac{1}{\sin \theta}\partial_r V_\varphi &= \tfrac{1}{2} \left({\cal H}_I \partial_\theta {\cal H}^I - {\cal H}^I\partial_\theta {\cal H}_I \right) \;. \label{eq:STU_KKvector2}
	\end{align}
	Since solutions should be asymptotically flat, we must require that 
$\partial_{[\mu}V_{\nu]} \longrightarrow 0$ as $r \longrightarrow \infty$. We will come back to this shortly. 
	Single-centred harmonic functions in oblate spheroidal coordinates can be written as
	\begin{align*}
		{\cal H}^I &= h^I + \frac{p^Ir + m^I \alpha\cos\theta}{R} 
\;,\\
		{\cal H}_I &=  h_I + \frac{q_Ir + m_I \alpha\cos\theta}{R} \;,\\
	\end{align*}
	where $R = r^2 + \alpha^2\cos^2\theta$.
It is understood that $(h^I,h_I,m^I,m_I,p^I,q_I)$ are all independent integration constants.	
While $h^I, h_I$ determine the values of the scalars
at infinity and $p^I,q_I$ are the magnetic and electric
charges, $m^I,m_I$ are the dipole momenta \cite{Behrndt:1997ny}. 
In \cite{Behrndt:1997ny} a restricted class of harmonic functions was considered, which corresponds to switching off half of the integration constants appearing in the expressions above. This restricted class of solutions was taken in order to satisfy the condition that the field strength of the $U(1)$ connection vanishes. In our formalism it is clear that we do not need impose this condition to produce solutions. 
	
	Integrating the equations (\ref{eq:STU_KKvector1}) and (\ref{eq:STU_KKvector2}) we find an explicit model independent expression for the only non-zero component of the KK-vector 
	\begin{align}
		V_{\varphi} = \, &\frac{1}{2}(h_I p^I - h^Iq_I)\cos \theta \left(\frac{r^2 + \alpha^2}{R} \right) \notag\\
		&+ \frac{\alpha}{2}(m_I h^I - m^I h_I) \sin^2 \theta \left(\frac{r}{R} \right) \notag \\
		&+ \frac{\alpha}{4}(m_I p^I - m^I q_I) \sin^2 \theta \left( \frac{1}{R} \right) + C \;, \label{eq:KK_vec}
	\end{align}
	where $C$ is an arbitrary constant. 
We observe that all three independent symplectic constructions
of the vectors $(h^I, h_I)$, $(p^I, q_I)$ and $(m^I, m_I)$ of
integration constants appear in this expression. The term in 
the second line is the angular momentum of the black hole, while
$n=\frac{1}{2} (h_I p^I - h^I q_I)$ is the NUT charge, as can be
seen by comparison with \cite{Bellorin:2006xr,Bossard:2008sw}. 
The term in the third
line does not carry a particular name, but is known to occur in 
rotating solutions \cite{Bellorin:2006xr}.
For static solutions all these
terms are absent, which beside $m^I=m_I=0$ imposes the constraint
$h_I p^I - h^I q_I=0$ on the integration constants.
Note that upon imposing this condition the KK-vector reduces to 
	\begin{equation}
		V_{\varphi} = \frac{\alpha \sin^2 \theta}{R} \left[ \tfrac{1}{2}\Big(m_I h^I - m^I h_I\Big) r + \tfrac{1}{4}\Big( m_I p^I - m^I q_I \Big) \right] \;. \label{V}
	\end{equation}
	Since this is proportional to $\alpha$ it will vanish in the static 
limit. 
In the general case $V_\varphi$ does note vanish for $r\rightarrow \infty$ 
unless we impose $h_Ip^I-h^Ip_I=0$ (and $C=0$). However, since 
the field strength $\partial_{[\mu} V_{\nu]}$ goes to zero, such a term
could be eliminated by a coordinate transformation. 
	In addition to requiring the KK-vector to vanish asymptotically, we also need to ensure the KK-scalar behaves appropriately, i.e.\ $e^\phi \longrightarrow 1$ as $r \longrightarrow \infty$. This will place one more restriction on the integration constants $(h^I, h_I)$. Since the KK-scalar is a model dependent field we will need to look at specific examples if we wish to write this constraint explicitly.


The formula for the ADM mass for axially symmetric solutions is given by
	\begin{align*}
		16 \pi M_{\text{ADM}} &= 2 \oint_{S^2_\infty} d^2 \Sigma^r e^{-\phi} \partial_r \phi \;. 
	\end{align*}	
	Expanding in descending orders of $r$ we have
	\begin{align*}
		d^2 \Sigma^r &= (r^2 + {\cal O}\left(r\right)) \sin \theta \; d\theta d\varphi \;,
		\qquad
		e^{-\phi} = 1 + {\cal O}\left(\frac{1}{r}\right) \;.
	\end{align*}
	Computing the ADM mass one finds a particularly simple dependence on the Hesse potential
	\begin{align*}
		M_{\text{ADM}} 
		&= - \lim_{r \to \infty} r^2 \partial_r \tilde{H} \;.
	\end{align*}	
	We would now like to investigate the relation between the mass and central charge. For solutions with vanishing NUT charge one has $ r^2 q^a \Omega_{ab} \partial_r q^b \to 0$  asymptotically, which implies that $r^2 q_a \Omega^{ab} \partial_r q_b \to 0$ asymptotically. We can then write the mass as	
	\begin{align}
		 M_{\text{ADM}} &= \lim_{r \to \infty} r^2 \Big( q^a - iH\Omega^{ab}q_b \Big) \partial_r q_a \;, \notag \\
		 					&= \lim_{r \to \infty} \left| X^I q_I - F_I p^I \right| = \lim_{r \to \infty} |{\mathcal Z}| \;.
	\end{align}
	This confirms that these solutions are BPS.


Before we enter into a discussion of specific models, we need to make 
a few comments about this class of rotating solutions. It contains 
the rotating supersymmetric solutions of \cite{Behrndt:1997ny},
which are not black holes but have naked singularities. As is well known,
for rotating four-dimensional solutions the extremality bound is higher
than the supersymmetric mass bound, so that rotating supersymmetric
solutions are necessarily singular. 
Besides the ring singularity at $r=0$, a non-vanishing
NUT charge can introduce further singularities \cite{Bossard:2008sw}. We also
remark that time-independence might imply further constraints
on the allowed charges \cite{Denef:2000nb,VandenBleeken:2011ib}.
Due to such constraints and the presence of naked singularities, 
the physical relevance of these rotating solutions is not 
immediately clear, in contrast to 
the static solutions to be considered later.
For us they are interesting for 
technical reasons, because they show how rotating solutions can be 
obtained within the framework of dimensional reduction over time.
To obtain physically relevant rotating solutions without naked
singularity our method needs to be extended to solutions which 
take values along non-isotropic submanifolds. This is similar to
the problem of deforming static extremal into non-extremal black holes,
and both problems will be addressed in future work. We conclude
this section by giving the explicit solution for the STU model.

\subsubsection{The $STU$ model}
	
	For the STU model we can find solutions explicitly in closed form. The model is characterised by the prepotential
	\[
		F = -\frac{Y^1 Y^2 Y^3}{Y^0} \;.
	\]
The name $STU$-models derives from the conventional notation 
$S,T,U=\frac{Y^i}{Y^0}$, $i=1,2,3$ for the physical scalars.
The corresponding Hesse potential is given in terms of the imaginary parts of $Y^I, F_I$ by
	\begin{equation}
		H = -2 \sqrt{ -(u^Iv_I)^2 + d_{ABC}u^Bu^Cd^{ADE}v_Dv_e + 4u^0v_1v_2v_3 - 4v_0u^1u^2u^3 } \;, \label{eq:STU_H}
	\end{equation}
	where $d_{ABC} = |\epsilon_{ABC}|$. A detailed derivation of this expression is given in appendix A.1.

	Rotating isotropic solutions to this model correspond to taking $\frac{1}{H}u^I = {\cal H}^I$ and $\frac{1}{H}v_I  = {\cal H}_I$. Using the expression $e^\phi = -2H$ we can write the KK-scalar for the STU model explicitly in terms of harmonic functions
	\[
		e^{-\phi} = \sqrt{ -({\cal H}^I{\cal H}_I)^2 + d_{ABC}{\cal H}^B {\cal H}^C d^{ADE} {\cal H}_D {\cal H}_e + 4{\cal H}^0 {\cal H}_1 {\cal H}_2 {\cal H}_3 - 4{\cal H}_0 {\cal H}^1 {\cal H}^2 {\cal H}^3 } \;. 		
	\]

	In order that the solution is asymptotically Minkowski space we must impose a constraint on the integration constants 
	\[
		-(h^Ih_I)^2 + d_{ABC}h^Bh^Cd^{ADE}h_Dh_e + 4h^0h_1h_2h_3 - 4h_0h^1h^2h^3  = 1 \;.
	\]
	At first glance it also appears that the KK-vector (\ref{eq:KK_vec}) will not vanish asymptotically, as is required for Minkowski space. However, since the  field strength of the KK-vector vanishes asymptotically we can make a change of coordinates so that spacetime is Minkowski. 
	
	For completeness, let us remark on the remaining four-dimensional fields for this solution. The original complex scalar fields are given by
	\begin{equation}
		X^I = e^{-\frac{\phi}{2}} Y^I \;, \hspace{3em} \bar{X}{}^I = e^{-\frac{\phi}{2}} \bar{Y}{}^I \;, \label{eq:STUrot_original_scalars}
	\end{equation}
	where $Y^I$ are given in terms of $u^I,v_I$ through 
	\begin{align}
		&Y^0 = \frac{1}{U + \bar{U}} \left( 2u^3 + i2u^0 \bar{U}\right) \;, & &Y^1 = \frac{1}{U + \bar{U}} \left( -2v_2 + i2 u^1 \bar{U} \right)\;, \notag\\
		&Y^2 = \frac{1}{U + \bar{U}}\left( -2v_1 +i2u^2 \bar{U} \right) \;, & &Y^3 = iUY^0 \;, \label{eq:STU_scalars}
	\end{align}
	with
	\begin{align}
		U =\, &i\frac{v_0u^0 + v_1u^1 + v_2 u^2 - v_3 u^3}{2\left(v_3u^0 + u^1 v^1\right)} \notag \\
		&\pm \sqrt{\frac{v_1v_2 - v_0u^3}{v_3u^0 + u^1u^2} - \frac{\left(v_0u^0 + v_1u^1 + v_2 u^2 - v_3 u^3 \right)^2}{4(v_3u^0 + u^1u^2)^2}} \;. \label{eq:STU_U}
	\end{align}
	These expressions have been adapted from similar expressions derived in \cite{Behrndt:1996hu}. One can substitute $u^I = -\frac12 e^{\phi/2} {\cal H}^I$ and $v_I = -\frac12 e^{\phi/2} {\cal H}_I$ to obtain the solution explicitly in terms of harmonic functions. The gauge fields are given by the expressions (\ref{eq:gauge1}),(\ref{eq:gauge2}).

\section{Static Solutions}

\subsection{General discussion}

When we impose that solutions are static and not only stationary, the
isotropic ansatz provides us with extremal black hole solutions. This
class is therefore of imminent physical importance.
	Static backgrounds are characterised by a vanishing KK-vector $V_\mu = 0$, which in dualised fields corresponds to
	\[
		\frac{1}{2H} \left( \partial_\mu \tilde{\phi} + 2\hat{q}^a \Omega_{ab} \partial_\mu \hat{q}^b \right) = 0 \;.		
	\]
	To obtain static solutions we will impose precisely the same isotropic ansatz as for stationary solutions, but in order to link to previous work we will reverse the order in which we apply the two parts of the ansatz. We first impose only the second part of the isoptropic ansatz (\ref{eq:ansatz2}), which in this case is simply 
	\begin{equation}
		q^a\Omega_{ab}\partial_\mu q^b = \pm q^a\Omega_{ab}\partial_\mu \hat{q}^b = \tfrac{1}{2}\left(\partial_\mu \tilde{\phi} + 2\hat{q}^a \Omega_{ab} \partial_\mu \hat{q}^b \right)  = 0 \;. \label{eq:static_ansatz1}
	\end{equation}
	It is then clear that the equations of motion simplify considerably. Only the first line of each equation is relevant, and we are have left with
	\begin{align}
		&\nabla^\mu \left[\tilde{H}_{ab}\partial_\mu q^b \right] - \tfrac{1}{2}\partial_a \tilde{H}_{bc} \left(\partial_\mu q^b \partial^\mu q^c - \partial_\mu \hat{q}^b \partial^\mu \hat{q}^c \right) = 0 \;, \label{eq:eom_static1} \\
		&\nabla^\mu \left[\tilde{H}_{ab}\partial_\mu \hat{q}^b \right] = 0 \;,	\label{eq:eom_static2} \\
		&\tilde{H}_{ab} \left(\partial_\mu q^a \partial_\nu q^b - \partial_\mu \hat{q}^a \partial_\nu \hat{q}^b \right)  \label{eq:eom_static3} = -\tfrac{1}{2} \tilde{R}_{3\mu\nu} \;.
	\end{align}	
	The equation of motion corresponding to the KK-vector is clearly solved automatically. The effective action for these equations is given by the first line of (\ref{eq:Lag_Hess})
	\[
			\tilde{{\cal L}}_{3} \sim  - \tfrac{1}{2} \tilde{R}_3 - \tilde{H}_{ab} \left(\partial_\mu q^a \partial^\mu q^b - \partial_\mu \hat{q}^a \partial^\mu \hat{q}^b \right)  \;. 
	\]

	The equations of motion (\ref{eq:eom_static1}) (\ref{eq:eom_static2}) and (\ref{eq:eom_static3}) take precisely the same form as when one reduces five-dimensional vector-multiplets over a timelike dimension in static, purely electric backgrounds. Both isotropic and non-isotropic solutions have been found in this case, and can be shown to lift to electrically charged extremal black holes \cite{Mohaupt:2009iq} and non-extremal black holes respectively \cite{Mohaupt:2010fk}. In order to obtain non-isotropic solutions one must modify (\ref{eq:ansatz1}), the part of our ansatz that relates $\partial_\mu q$ and $\partial_\mu \hat{q}$, by a universal `non-extremality' factor. In this case the three-dimensional spacetime metric is no longer flat but conformally flat. The machinery for producing these non-isotropic solutions takes a slightly different form than in the isotropic case, and for that reason we will not consider these solutions in this paper. We remark that is possible to use the techniques established in \cite{Mohaupt:2010fk} to produce non-isotropic solutions which lift to non-extremal black holes in four-dimensions, which we have found for particular models, but we leave a detailed discussion of this topic to future work. 
	
	In order to produce isotropic solutions to these equations of motion in flat three-dimensional backgrounds we must again impose the ansatz
	\begin{equation}
		\partial_\mu q^a = \pm \partial_\mu \hat{q}^a \;. \label{eq:static_ansatz2}
	\end{equation}
	It is clear by inspection that in this case all equations of motion reduce to the Laplace equation for the dual coordinates
	\[
		\Delta q_a = 0 \;.
	\]
	In this case condition (\ref{eq:static_ansatz1}) places one constraint on the integration constants of $q_a$.


	The formula for the ADM mass for is given by
	\begin{align*}
		16 \pi M_{\text{ADM}} &= 2 \oint_{S^2_\infty} d^2 \Sigma^\mu e^{-\phi} \partial_\mu \phi \;. 
	\end{align*}	
	Since $e^\phi \to 1$ at spatial infinity we  can write this as
	\begin{align*}
		 M_{\text{ADM}} &= -\frac{1}{4\pi} \oint_{S^2_\infty} d^2 \Sigma^\mu \partial_\mu \tilde{H} \;,
	\end{align*}
	Using the fact that the NUT charge vanishes $q^a \Omega_{ab} \partial_\mu q^b = 0$, which implies that  $q_a \Omega^{ab} \partial_\mu q_b = 0$, we can write this as
	\begin{align}
		M_{\text{ADM}} &= \frac{1}{4\pi} \oint_{S^2_\infty} d^2 \Sigma^\mu \left(q^a  - i H \Omega^{ab} q_b \right) \partial_\mu q_a \;, \notag \\
				&=  \frac{1}{4\pi} \oint_{S^2_\infty} d^2 \Sigma^\mu \left| X^I \partial_\mu {\cal H}_I - F_I \partial_\mu {\cal H}^I \right| = |{\cal Z}_\infty| \;. \label{eq:ADMmass}
	\end{align}
	These extremal black hole solutions therefore satisfy the BPS bound.	


\subsection{Examples of extremal black hole solutions}

	We will now consider explicit solutions to the equations of motion in static backgrounds. We impose the ans\"atze (\ref{eq:static_ansatz1}) and (\ref{eq:static_ansatz2}) and solutions are again given by harmonic functions, but in this case they are not bound by any symmetry constraints. Solutions correspond to extremal black holes in four-dimensions in the sense they have finite horizons, are asymptotically Minkowski, and saturate a bound on the mass and charge.

	We will first consider a class of extremal black hole solutions of the STU model that are obtained by taking the static limit of the rotating solutions discussed in the previous section. We will then present axion-free solutions to a wider class of models which have prepotentials of the form $F(Y) = \frac{f(Y^1,\ldots,Y^n)}{Y^0}$. This class of models includes those that have a `very special' form, where $f(Y^1,\ldots, Y^n)$ is a homogeneous cubic polynomial. Such models can which be obtained by the dimensional reduction of five-dimensional theories. While axion-free solutions for very special prepotentials are well known 
\cite{Behrndt:1996jn}, our derivation shows that to obtain solutions it is 
enough to assume that $f$ is homogeneous, and so we can obtain axion-free
solutions for a larger class of prepotentials.

We end by giving explicit solutions to models where
$f=STU + aU^3$. This is a deformation of the $STU$-model which
is still of the very special form, but the target space is no longer
symmetric. The model with $a=\frac{1}{3}$ corresponds to a particular
Calabi-Yau compactification and its heterotic dual 
\cite{Morrison:1996pp,Louis:1996mt}.


\subsubsection{The $STU$ model}

	We first consider the static limit of the rotating solutions found in the previous section. This will give us extremal black hole solutions to the STU model. Taking the static limit amounts to setting $\alpha \to 0$ and imposing the constraint 
	\[
		h^Iq_I - h_I p^I = 0 \;,
	\]
	which ensures the KK-vector vanish identically. The dipole momenta $m^I, m_I$ completely vanish from the solution, along with angular momentum and NUT charge, and we are left with a spherically symmetric configuration.
	The expression for the KK-scalar remains unchanged, and we obtain the solution
	\[
		e^{-\phi} = \sqrt{ -({\cal H}^I {\cal H}_I)^2 + (d_{ABC}{\cal H}^B {\cal H}^C d^{ADE}{\cal H}_D {\cal H}_e) + 4{\cal H}^0 {\cal H}_1 {\cal H}_2 {\cal H}_3 - 4{\cal H}_0 {\cal H}^1 {\cal H}^2 {\cal H}^3 } \;,		
	\]
	\[
			g_{\mu\nu} = \delta_{\mu\nu} \;, \hspace{3em} V_\mu = 0 \;.
	\]
	where the harmonic functions are given by
	\begin{align*}
		{\cal H}^I &= h^I + \frac{p^I}{r}  \;, \\
		{\cal H}_I &= h_I + \frac{q_I}{r}  \;,
	\end{align*}
	The asymptotic integration constants $h^I, h_I$ satisfy the two constraints
	\begin{align*}
		-(h^Ih_I)^2 + (d_{ABC}h^Bh^Cd^{ADE}h_Dh_e) + 4h^0h_1h_2h_3 - 4h_0h^1h^2h^3 &= 1 \;, \\
		h_I p^I - h^Iq_I &= 0 \;.
	\end{align*}
	Like in the case for rotating solutions, using the expression 
(\ref{eq:phi_H}) we can write $u^I, v_I$ explicitly in terms of harmonic functions by 
	\begin{equation}
		u^I = -\tfrac{1}{2}e^\phi {\cal H}^I \;, \hspace{3em} v_I = -\tfrac{1}{2}e^\phi {\cal H}_I.
		\label{eq:STU_vu}
	\end{equation}
	The original four-dimensional scalar fields are given by
	\[
		X^I = e^{-\frac{\phi}{2}} Y^I \;, 
	\]
	where $Y^I$ are given in terms of harmonic functions through (\ref{eq:STU_scalars}) and (\ref{eq:STU_vu}). Again, the gauge fields are given by the expressions (\ref{eq:gauge1}),(\ref{eq:gauge2}).

	The above extremal black hole solutions of the STU model are spherically symmetric as they were obtained by taking the static limit of axially-symmetric rotating solutions, but this need not be the case in general. If we do not impose any symmetry constraints on spacetime then we will obtain the same expressions for the four-dimensional fields, but with the harmonic functions which are completely general. Multi-centered black hole solutions with centers
are at $x_\alpha$ correspond to the choice 
	\begin{align*}
		{\cal H}^I &= h^I + \sum_\alpha \frac{p^I_\alpha}{|x - x_\alpha|} 
\;,\\
		{\cal H}_I &= h_I + \sum_\alpha \frac{q_{I\alpha}}{|x - x_\alpha|} \;.
	\end{align*}

\subsubsection{Models of the form $F = \frac{f(Y^1, \ldots, Y^n)}{Y^0}$}

	A class of models for which we can find explicit extremal black hole 
solutions are those where the prepotential takes the form
	\[
		F(Y) = \frac{f(Y^1, \ldots, Y^n)}{Y^0} \;,
	\]
	where $f$ is real when evaluated on real fields. Since $F$ is a homogeneous function of degree 2 it follows that $f$ is a homogeneous function of degree 3. If $f$ is in particular a cubic polynomial 
$f=C_{ABC} Y^A Y^B Y^C$ with real $C_{ABC}$, 
then this is of the `very special' form which 
derives from five-dimensional supergravity by reduction.

We will consider a restricted set of solutions that are characterised by the requirement that $Y^A$ are purely  imaginary and $Y^0$ is purely real, which 
implies that the four-dimensional scalars $Z^A=Y^A/Y^0$  are purely 
imaginary. For very special prepotentials, where the real part of $Z^A$
corresponds to a five-dimensional gauge potential, this means that
such solutions are `axion-free.' For $f=C_{ABC} Y^A Y^B Y^C$ it follows
that $F_0$ is imaginary while $F_A$ are real. If we replace 
$C_{ABC} Y^A Y^B Y^C$ by a general homogeneous function $f$ of degree three
this remains true only if we impose that $f$ is real when evaluated on 
real fields $Y^A$ (and, by homogeneity, imaginary when evaluated on 
imaginary fields $Y^A$). Therefore we , we impose this condition
in the following, and `axion-free solutions' are characterized by the 
consistent reality condition on the fields which impose that 
$Y^0$ and $F_A$ are purely real while $Y^A$ and $F_0$ are purely 
imaginary. In terms of real variables this corresponds to imposing that
\begin{equation}
x^1 = \ldots = x^n = y_0 = 0 \;,
		\label{eq:ax_free_condition}
\end{equation}
which defines a particular submanifold of the scalar manifold.
In terms of dual coordinates (\ref{eq:ax_free_condition}) 
is equivalent to 
\begin{equation}
v_1 = \ldots = v_n = u^0 = 0 \;.
\end{equation}

	With the above assumptions we can write $Y^I,F_I$ in terms of the dual
real coordinates as
	\begin{equation}
	\label{eq:ax_free_f}
	\begin{aligned}
		Y^0 &= \lambda \;, & F_0 &= i v_0 \;,  \\ 
		Y^A &= iu^A \;,  & F_A &= -\frac{f_A(u^1, \ldots, u^n)}{\lambda} \;, 	
	\end{aligned}
	\end{equation}
	where 
	\[
		\lambda = -\sqrt{\frac{f(u^1, \ldots, u^n)}{v_0}} \;,
	\]
	and $f_A = \frac{\partial f}{\partial Y^A}$.
	Using (\ref{eq:phi_K}) and (\ref{eq:phi_H}) we obtain expressions for the KK-scalar and Hesse potential (evaluated on axion-free configurations) 
	\[
		e^{\phi} = -2H = -i(Y^I \bar{F}_I - F_I \bar{Y}^I) = 8 \sqrt{ v_0 f(u^1, \ldots, u^n)} \;.
	\]
	The real parts of $Y^I, F_I$ can be read off from (\ref{eq:ax_free_f}) as
	\begin{equation}
		x^0 = \lambda  \;, \hspace{3em} y_A = \frac{f_A(u^1, \ldots, u^n)}{\lambda} \;. \notag 
	\end{equation}
	This amounts to solving the generalised stabilisation equations, and is the reason why we can find solutions explicitly in closed form.
	
	Solutions to these models are given in terms of harmonic functions by 
	\[
		e^{-\phi} = \sqrt{ 4{\cal H}_0 f({\cal H}^1, \ldots, {\cal H}^n) } \;,
	\]
	\[
			g_{\mu\nu} = \delta_{\mu\nu} \;, \hspace{3em} V_\mu = 0 \;.
	\]
	The harmonic functions are given by
	\[
		{\cal H}_0 = h_0 + \sum_\alpha \frac{q_{0\alpha}}{|x - x_\alpha|} \;, \hspace{3em}	{\cal H}^A = h^A + \sum_\alpha \frac{p^A_\alpha}{|x - x_\alpha|} \;,
	\]
	with ${\cal H}_A = {\cal H}^0 = 0$.

	The asymptotic integration constants $h^I, h_I$ must satisfy only one constraint
	\[
		4h_0 f(h^1, \ldots, h^n) = 1 \;.
	\]
	We can write $v_0, u^A$ explicitly in terms of harmonic functions by 
	\begin{equation}
		v_0 = -\tfrac{1}{2}e^\phi {\cal H}_0 \;, \hspace{3em} u^A = -\tfrac{1}{2}e^\phi {\cal H}^A \;.
		\label{eq:f_vu}
	\end{equation}
	The original four-dimensional scalar fields are given by
	\[
		X^I = e^{-\frac{\phi}{2}} Y^I \;, 
	\]
	which can be written in terms of harmonic functions using (\ref{eq:ax_free_f}) and (\ref{eq:f_vu}). The gauge fields are given by the expressions (\ref{eq:gauge1}),(\ref{eq:gauge2}).

\subsubsection{The $STU + aU^3$ model}

We now turn to a specific one-parameter family of models of the form $F = \frac{f(Y^1, Y^2, Y^3)}{Y^0}$, which are characterised by the prepotential
\[
	F(Y) = -\frac{Y^1Y^2Y^3 + a(Y^1)^3}{Y^0} \;.
\]
This is a deformation of the $STU$-model where the target space is no
longer symmetric. 
Specialising to solutions with $x^1 = x^2 = x^3 = 0$ and $y_0 = 0$ we have  
	\begin{equation}
	\label{eq:STU_U3_YF}
	\begin{aligned}
		Y^0 &= \lambda \;, & \qquad F_0 &= i v_0 \;,  \\ 
		Y^1 &= iu^1 \;,  & F_1 &= \frac{ u^2 u^3 + 3a (u^1)^2}{\lambda} \;,  \\
		Y^2 &= iu^2 \;,  & F_2 &= \frac{ u^1 u^3}{\lambda} \;,  \\
		Y^3 &= iu^3 \;,  & F_3 &= \frac{ u^1 u^2}{\lambda} \;,  
	\end{aligned}
	\end{equation}
	where 
	\[
		\lambda = -\sqrt{-\frac{u^1 u^2 u^3 + a(u^1)^3}{v_0}} \;.
	\]

Solutions are given in terms of harmonic functions by
	\[
		e^{-\phi} = \sqrt{ -4{\cal H}_0 ({\cal H}^1{\cal H}^2{\cal H}^3 + a ({\cal H}^1)^3) } \;,
	\]
	\[
			g_{\mu\nu} = \delta_{\mu\nu} \;, \hspace{3em} V_\mu = 0 \;,
	\]
	where the harmonic functions are again defined to be
	\[
		{\cal H}_0 = h_0 + \sum_\alpha \frac{q_{0\alpha}}{|x - x_\alpha|} \;, \hspace{3em}	{\cal H}^A = h^A + \sum_\alpha \frac{p^A_\alpha}{|x - x_\alpha|} \;,
	\]
	with ${\cal H}_A = {\cal H}^0 = 0$.
	The asymptotic integration constants $h^I, h_I$ must satisfy the constraint
	\[
		-4h_0 (h^1 h^2 h^3 + a {h^1}^3) = 1 \;.
	\]
	We can write $v_0, u^A$ explicitly in terms of harmonic functions by 
	\begin{equation}
		v_0 = -\tfrac{1}{2}e^\phi {\cal H}_0 \;, \hspace{3em} u^A = -\tfrac{1}{2}e^\phi {\cal H}^A \;.
		\label{eq:STU_U3_VU}
	\end{equation}
	The original four-dimensional scalar fields can be determined through the expressions
	\[
		X^I = e^{-\frac{\phi}{2}} Y^I \;, 
	\]
	which one can write explicitly in terms of harmonic functions using (\ref{eq:STU_U3_YF}) and (\ref{eq:STU_U3_VU}).
	The gauge fields are given by the expressions (\ref{eq:gauge1}),(\ref{eq:gauge2}).

\subsection{Field rotations and non-BPS solutions}

Four-dimensional extremal non-BPS have been studied in the
past \cite{Khuri:1995xq,Ortin:1996bz,Ortin:1997yn}, and more recently
there has been increased interest in this topic,
starting from \cite{Goldstein:2005hq,Tripathy:2005qp,Kallosh:2005ax}.
As in the five-dimensional case \cite{Mohaupt:2009iq}, the ansatz 
$\partial_\mu q^a = \pm \partial_\mu \hat{q}^a$ with a universal sign
does not necessarily exhaust all solutions. 
To obtain further solutions we can adapt
the observation that new solutions can be generated by flipping signs
of charges \cite{Ortin:1996bz}, or, more generally, by `rotating charges' 
\cite{Ceresole:2007wx,LopesCardoso:2007ky}. 
BPS solutions correspond to particular combinations of signs,
while other choices lead to non-BPS solutions.

As we have seen above the ansatz $\partial_\mu q^a 
= \pm \partial_\mu \hat{q}^a$ leads to BPS solutions. For static 
solutions we can use the same generalization of the ansatz 
as in five-dimensions \cite{Mohaupt:2009iq} and introduce a constant 
field rotation matrix 
	\begin{equation}
		\partial_\mu q^a = R^a_{\;\;b} \partial_\mu \hat{q}^b \;.
\label{eq:static_ansatz2_rot}
	\end{equation}
This is the analogue of `rotating charges' in our framework. 
By inspection of the field equations, we find
that this ansatz only works if the following compatibility condition
between the scalar metric and the field rotation matrix holds
	\begin{equation}
		\tilde{H}_{ab}R^a_{\;\;c}R^b_{\;\;d} = \tilde{H}_{cd} \;. 
\label{eq:charge_rot}
	\end{equation}
If this condition is satisfied, then the solution for the dual scalar
fields $q_a, \hat{q}_a$ is again given by harmonic functions, but now
the harmonic functions for $q_a$ are related to those for $\hat{q}_a$
through the constant matrix $R_a^{\;\;b}$, which is the transposed
of the inverse of $R^a_{\;\;b}$:
\[
\partial_\mu q_a = \partial_\mu {\cal H}_a =
R_a^{\;\;b} \partial_\mu \hat{q}_b  = R_a^{\;\;b} \partial_\mu \hat{H}_b\;,\;\;\;
R^a_{\;\;b} R_a^{\;\;c} = \delta_b^c \;.
\]

 Equivalently, the relations
(\ref{eq:gauge1}) and (\ref{eq:gauge2}) between four-dimensional
scalars and gauge fields are modified by the presence of this matrix. 
Decomposing the field rotation matrix $R^{T,-1}$ into blocks
	\[
		R^{T,-1} = \left( \begin{array}{cc}
A & B \\
C & D \\
\end{array} \right) \;,
	\]
	the expressions for the gauge fields become
	\begin{align}
 \partial_\mu (e^{\phi/2} (X^J + \bar{X}^J))A_J^{\;\;\;I} + \partial_\mu (e^{\phi/2}(F_J + \bar{F}_J )) C^{JI} &= \pm (F^{I|+}_{0\mu} + F^{I|-}_{0\mu}) \;, \label{eq:gauge1_rot} \\
 \partial_\mu (e^{\phi/2} (X^J + \bar{X}^J))B_{JI} +\partial_\mu (e^{\phi/2}(F_J + \bar{F}_J )) D^J_{\;\;\;I} &= \pm (G^+_{I|0\mu} + G^-_{I|0\mu}) \;. \label{eq:gauge2_rot}
\end{align}
	In particular, the electric and magnetic charges appear rotated relative to the solutions of the scalar fields. Note that in general not only the
charges but also the asymptotic behaviour of solutions changes 
\cite{Galli:2011fq}. This is necessary in order to avoid introducing
naked singularities.\footnote{We thank the authors of \cite{Galli:2011fq}
for bringing this to our attention.}

The presence of a non-trivial field rotation matrix also modifies the ADM mass 
(\ref{eq:ADMmass}):
	\[
		M_{\text{ADM}} = \frac{1}{4\pi} \oint_{S^2_\infty} d^2 \Sigma^\mu \left| X^I \left( A_{I}^{\;\;J} \partial_\mu \hat{\cal H}_J + B_{IJ} \partial_\mu 
\hat{\cal H}^J \right) - F_I \left( C^{IJ} \partial_\mu  \hat{\cal H}_J 
+ D^{I}_{\;\;J} 
\partial_\mu \hat{\cal H}^J \right)  \right| \;.
	\]
This makes it manifest that such solutions are not BPS. 
Note that we saw above that the $R=\pm \mbox{Id}$ 
leads precisely to the relation between four-dimensional scalars and 
gauge fields which is implied by the BPS condition. 


Since a field rotation matrix only provides a solution if the 
compatibility condition (\ref{eq:charge_rot}) is satisfied, it is 
in general not clear that non-BPS solutions can be obtained by this
ansatz. For symmetric spaces non-BPS solutions can be obtained
in a systematic way using group-theoretical methods 
\cite{Gaiotto:2007ag,Bossard:2009we}. For non-symmetric target spaces
these methods do not apply, and therefore
it is interesting to ask under which conditions one can guarantee 
the existence of a non-trivial field rotation matrix which satisfies
(\ref{eq:charge_rot}). 

Geometrically, this is equivalent to the problem of identifying 
totally geodesic, totally isotropic submanifolds of the scalar 
target space. We have seen that for $c$-map spaces there is a
universal solution, given by the ansatz $\partial_\mu q^a = \pm 
\partial_\mu \hat{q}^a$, which corresponds to BPS solutions. Finding
non-BPS solutions amounts to finding further such submanifolds, 
which correspond to the non-BPS branches that one can 
identify in symmetric target spaces by group theoretical methods.

In the following section we establish that a non-trivial field
rotation matrix exists for non-axionic solutions of models
with a prepotential of the form $F = \frac{f(Y^1, \ldots, Y^n)}{Y^0}$
where $f$ is real when evaluated on real fields,
i.e. for the class of examples considered above. Before we turn
to the details, we remark that we do not only need to impose
a condition on the model (i.e. on the form of the prepotential),
but also on the field configurations, by restricting to axion-free
solutions. This corresponds to restricting to  lower-dimensional
submanifolds of the scalar manifold. If no such restriction is
imposed, the 
compatibility condition (\ref{eq:charge_rot}) implies
that the field rotation matrix acts by an isometry. Requiring 
the existence of such an isometry imposes a condition on the prepotential.
By restricting to field configurations which are axion free, the
compatibility condition (\ref{eq:charge_rot}) becomes less restrictive
and we can establish the existence of a field rotation matrix under
much milder assumptions on the form of the prepotential. But the resulting 
totally geodesic, totally isotropic submanifold corresponding to the
axion-free non-BPS solution is of lower dimension than the submanifold
corresponding to BPS solutions, which has maximal dimension. It would
be interesting to clarify whether this is a generic feature of non-BPS
solutions in models with non-symmetric target spaces.

Finally, we mention that in the rotating case one cannot simply adapt the isotropic ansatz in the same way when a field rotation matrix is available, as this no longer produces a solution to the equations of motion. In order to produce non-BPS rotating solutions one needs to relax the condition that three-dimensional metric is flat. We will not consider such solutions in this paper, and leave the investigation of such solutions to future work.

\subsubsection{Non-BPS solutions to $F = \frac{f(Y^1, \ldots, Y^n)}{Y^0}$ models}

For models with prepotentials of the form $F = \frac{f(Y^1, \ldots, Y^n)}{Y^0}$, where $f$ is real when evaluated on real fields, there always exists a  non-trivial field rotation matrix for solutions satisfying the conditions (\ref{eq:ax_free_condition}).
For the remainder of this section we will focus on the specific case where $n = 3$, but the solutions can be extended to arbitrary $n \geq 1$ without loss of generality.

To see why a field rotation matrix always exists for this class of models we must analyze the matrix $\tilde{H}_{ab}$ in some detail. Firstly, one observes that the conditions (\ref{eq:ax_free_condition}) imply that the matrix $\tilde{H}_{ab}$ decomposes into
		\begin{equation}
			\tilde{H}_{ab} = \left( \begin{array}{c|cccc|ccc} 
			* & 0 & 0 & 0 & 0 & * & * & * \\ \hline
			0	& * & * & * & * & 0 & 0 & 0 \\
			0	& * & * & * & * & 0 & 0 & 0 \\ 
			0	& * & * & * & * & 0 & 0 & 0 \\
			0	& * & * & * & * & 0 & 0 & 0 \\ \hline
			* & 0 & 0 & 0 & 0 & * & * & * \\		
			* & 0 & 0 & 0 & 0 & * & * & * \\		
			* & 0 & 0 & 0 & 0 & * & * & *		
			\end{array} \right) \;, \label{eq:Htilde1}
		\end{equation}
		where a $*$ represents a possible non-zero entry. To see why this is the case, consider, for example, the matrix element $\tilde{H}_{10}$. Let us denote by $\sharp$ the restriction of solutions to (\ref{eq:ax_free_condition}). We can write $\tilde{H}_{10}$ as
		\begin{align*}
			\tilde{H}_{10}\Big|_{\sharp} &= \left(\frac{\partial}{\partial x^0} \frac{\partial \tilde{H}}{\partial x^1}\right)\Bigg|_{\sharp} \ \\
				&= \frac{\partial}{\partial x^0} \left(\frac{\partial \tilde{H}}{\partial x^1}\Bigg|_{\sharp} \right)  \\
				&= \frac{\partial}{\partial x^0} (0) = 0\;.
		\end{align*}
In the second line we used that the variable $x^0$ does not enter into the
axion-free condition $\sharp$, which amounts to setting other variables
to constant (zero) values. Therefore we can take the derivative with
respect to $x^0$ after imposing the axion free condition $\sharp$. 
		In the third line we used the fact that $\frac{\partial \tilde{H}}{\partial x^1} = \frac{-v_1}{H}$. This is valid irrespective of the 
condition $\sharp$ by definition of the dual coordinates. 
		The same argument is true for any matrix element containing one index in $\{0,5,6,7\}$ and one index in $\{1,2,3,4\}$. 
		
		When expressed in terms of $u^I, v_I$, the axion free 
ansatz (\ref{eq:ax_free_condition}) implies that $v_1 = v_2 = v_3  = u^0 =0$. 
Consequently, the corresponding harmonic functions vanish ${\cal H}_1 = {\cal H}_2 = {\cal H}_3  = {\cal H}^0 = 0$, and
the central $4 \times 4$ block appearing in (\ref{eq:Htilde1}) completely decouples from the equations of motion, and is of no relevance to the remaining discussion. 
		
		Actually, the matrix $\tilde{H}_{ab}$ decomposes even further. Using the formula for the Hesse potential (\ref{eq:axion_free_Hesse_xy}) for this class of solution, which is derived in appendix \ref{Hessepotential2}
one observes that $\tilde{H}_{ab}$ takes the more restrictive form
	\begin{equation}
		\tilde{H}_{ab} = \left( \begin{array}{c|cccc|ccc} 
		\frac{1}{4(x^0){}^2} & 0 & 0 & 0 & 0 & 0 & 0 & 0 \\ \hline
		0	& * & * & * & * & 0 & 0 & 0 \\
		0	& * & * & * & * & 0 & 0 & 0 \\ 
		0	& * & * & * & * & 0 & 0 & 0 \\
		0	& * & * & * & * & 0 & 0 & 0 \\ \hline
		0 & 0 & 0 & 0 & 0 & * & * & * \\		
		0 & 0 & 0 & 0 & 0 & * & * & * \\		
		0 & 0 & 0 & 0 & 0 & * & * & *		
		\end{array} \right) \;, \label{eq:Htilde2}
	\end{equation}
	where the entries in the bottom-right block depend only on $y_1, y_2, y_3$.
	
	For such solutions, these modes always admit a non-trivial field rotation matrix of the form
	\begin{equation}
		R^a_{\;\;b} = \pm \left( \begin{array}{c|ccc} 
		-1 & & 0 & \\ \hline
		\\
		0	& & {\mathbb I}_{2n + 1}& \\	
		\\
		\end{array} \right) \;.
		\label{eq:field_rot}
	\end{equation}
	One can therefore find non-BPS solutions to these models generically.
	
\subsubsection{Non-BPS solutions to $STU + aU^3$ model}

	Since this model falls into the category of $F = \frac{f(Y^1, Y^2, Y^3)}{Y^0}$ it admits the non-trivial field rotation matrix given by (\ref{eq:field_rot}), and we can obtain non-BPS solutions.

	The non-BPS solutions are given explicitly by 
	\[
		e^{-\phi} = \sqrt{ 4{\cal H}_0 ({\cal H}^1{\cal H}^2{\cal H}^3 + a ({\cal H}^1)^3) } \;,
	\]
	\[
			g_{\mu\nu} = \delta_{\mu\nu} \;, \hspace{3em} V_\mu = 0 \;,
	\]
	where the harmonic functions are again given by
	\[
		{\cal H}_0 = h_0 + \sum_\alpha \frac{q_{0\alpha}}{|x - x_\alpha|}
 \;, \hspace{3em}	{\cal H}^A = h^A + \sum_\alpha \frac{p^A_\alpha}{|x - x_\alpha|} \;,
	\]
	with ${\cal H}_A = {\cal H}^0 = 0$.	The asymptotic integration constants $h^I, h_I$ satisfy the constraint
	\[
		4h_0 (h^1 h^2 h^3 + a (h^1)^3) = 1 \;.
	\]
	We can write $v_0, u^A$ explicitly in terms of harmonic functions by 
	\begin{equation}
		v_0 = \tfrac{1}{2}e^\phi {\cal H}_0 \;, \hspace{3em} u^A = -\tfrac{1}{2}e^\phi {\cal H}^A \;.
		\label{eq:STU_U3_VU_nonBPS}
	\end{equation}
	The original four-dimensional scalar fields can be determined through the expressions
	\[
		X^I = e^{-\frac{\phi}{2}} Y^I \;, 
	\]
	which one can write explicitly in terms of harmonic functions using (\ref{eq:STU_U3_YF}) and (\ref{eq:STU_U3_VU_nonBPS}).
	The expressions for the gauge fields remain unchanged, and are given by (\ref{eq:gauge1}),(\ref{eq:gauge2}).

\section{Conclusions and Outlook}

In this paper we have shown how four-dimensional 
${\cal N}=2$ vector multiplets coupled to supergravity 
can be described in terms of a real formulation of special
K\"ahler geometry using the gauge equivalence with conformal
supergravity. Key technical points, which allowed us to preserve
symplectic covariance,  were to avoid $U(1)$ gauge fixing,
and the use of the degenerate metric obtained by integrating out
the auxiliary $U(1)$ gauge field. Geometrically this corresponds
to working on the Sasakian $S$ or the conical affine special K\"ahler
manifold $N$, and to use a horizontal lift for the metric. 
We expect that this formulation will be useful for
studying non-holomorphic corrections.

By dimensional reduction we have obtained a new
formulation of the supergravity $c$-map,
which is complementary to other existing formulations and 
offers new insights into the geometry as well as 
practical advantages for some types of problems. In our
formulation the local $c$-map comes very close to the 
Sasaki form of the  rigid $r$- 
and $c$-map, and of the local $r$-map. It is manifestly
symplectically invariant with respect to both vector and hypermultiplets,
it is completely formulated in terms of real variables, and it 
provides a simple and explicit expression for the quaternion-K\"ahler
metric in terms of the Hesse potential. We have
introduced a new geometrical object, a principal $U(1)$ bundle over
the quaternion-K\"ahler manifold, and work with the horizontal lift
of the metric to the total space of this bundle. 
We are currently investigating the deeper geometrical 
interpretation of our results and expect that this will be useful
for understanding the dynamics of hypermultiplets 
in string compactifications. One obvious question is the relation
of our construction to the hyper-K\"ahler cone and twistor space,
which could lead to a more complete picture of the $c$-map, 
hypermultiplets, and black hole and instanton solutions. 

When applied to the temporal version of the $c$-map, the
new parametrization makes it easy to find instanton
solutions which are restricted to totally isotropic
submanifolds. By dimensional lifting we have obtained extremal black holes and
over-extremal rotating solutions. Since the equations of motion
are reduced to decoupled harmonic equations, multi-centered 
solutions can be obtained as easily as single centered ones. The 
flexibility in choosing harmonic functions at the very end is 
an advantage of the method, which was further illustrated 
by constructing rotating solutions. 
Since the method does not rely
on Killing spinors it is not restricted to supersymmetric solutions. 
The black hole attractor equations
and other relations known from supersymmetric solutions are
derived from geometric properties of the scalar manifold and take
a manifestly symplectically covariant form. For static extremal 
solutions we thus obtain a full generalization of the previous results
on five-dimensional black holes \cite{Mohaupt:2009iq}. 

While the canonical version $\partial_\mu q^a = \pm \partial_\mu \hat{q}^a$
of the ansatz
always works and leads to BPS solutions, non-BPS solutions can be 
obtained if a non-trivial field rotation matrix exists, which 
must satisfy a compatibility condition with the metric. For non-symmetric
target spaces the existence of such a matrix is non-trivial, but we 
were able to show that it exists for axion-free solutions for a 
class of prepotentials, which contains the very special ones
as a subclass. An interesting future direction is to develop the
understanding of non-BPS solutions for non-symmetric target spaces.
Since symmetric spaces are contained in our formalism as special
cases, one promising strategy is to translate the group-theoretical
characterisations of BPS and non-BPS solutions into geometrical 
properties of totally geodesic, totally isotropic submanifolds 
and then to investigate whether these conditions have natural
generalizations for non-symmetric spaces.

Another direction 
is the generalization to non-extremal static black holes, which 
for the five-dimensional case was discussed in \cite{Mohaupt:2010fk},
and, more recently, in \cite{Meessen:2011bd}.
Deforming extremal into non-extremal solutions corresponds to
deforming isotropic into non-isotropic submanifolds. It is currently
not clear to us to which extent this can be done in a universal
way. However, specific examples suggest that our method can be 
generalized, and we plan to report on this in a future publication.
Non-extremal four-dimensional black holes in ${\cal N}=2$ supergravity 
have been recently 
discussed in \cite{Galli:2011fq} from a different though related
point of view. For ${\cal N}=4$ supergravity the full class of stationary
point-like solutions is known \cite{LozanoTellechea:1999my}.

We have also shown how rotating solutions can be obtained, and
recovered the known rotating supersymmetric solutions. In this
case the use of field rotation matrices to produce non-BPS
solutions requires to generalize the ansatz and to admit 
a curved three-dimensional base space. Moreover, these solutions
have naked singularities, and making non-singular will also
require to go beyond the isotropic ansatz considered in the 
second part of this paper.

\subsubsection*{Acknowledgments}

The work of T.M. is supported in part by STFC  
grants ST/G00062X/1 and ST/J000493/1.
The work of O.V. is supported by an STFC studentship and by DAAD. 
We thank Vicente Cort\'es, Bernard de Witand Dieter Van den Bleeken
for useful discussions. T.M. thanks the Department of Mathematics 
of the University of Hamburg and the SFB 676 for hospitality and
support during the final stages of this work. Finally, we thank
Patrick Meessen, Tomas Ortin, Jan Perz and C.S. Shahbazi for useful 
comments on the first version of this paper.

\appendix 

\section{Hesse potentials}

\subsection{Hesse potential for STU model}

In this section we derive the Hesse potential for the STU 
model. Due to the relation between the Hesse potential and BPS black hole 
entropy \cite{LopesCardoso:2006bg}, this is equivalent to solving
the attractor equations. However, the relation between Hesse potential
and prepotential is `off-shell', and does not require to impose a particular
background solution. Therefore we find it instructive to present the
derivation in a form where this is manifest. Technically we closely 
follow \cite{Behrndt:1996jn}, but instead of charges and horizon values of
fields we use fields without imposing supersymmetry or 
any of the field equations.

The STU model in special complex coordinates is characterised by the holomorphic prepotential
\[
	F(Y) = -\frac{Y^1 Y^2 Y^3}{Y^0} \;.
\]
Introducing the inhomogeneous coordinates $Z^A = Y^A/Y^0$ one can write the K\"ahler potential for the STU model as
\begin{equation}
	e^{\cal -K} = -i\left(Y^I \bar{F}_I - F_I \bar{Y}^I \right) = 8 Y^0 \bar{Y}^0 \text{Im}(Z^1)\text{Im}(Z^2)\text{Im}(Z^3) \;.
	\label{eq:KahlerPotential}
\end{equation}
Our strategy will be to write the individual fields $Y^0,Z^1,Z^2,Z^3$ in terms of $x^I = \text{Re}(Y^I)$ and $y_I = \text{Re}(F_I)$.

Firstly, by direct calculations one can show that
\[
	-\bar{Z}^2 \bar{Z}^3 = \frac{y_1 Z^1 + y_0}{x^0 Z^1 - x^1} \;, \hspace{2em} \bar{Z}^2 = \frac{x^2 Z^1 + y_3}{x^0 Z^1 - x^1} \;, \hspace{2em} \bar{Z}^3 = \frac{x^3 Z^1 + y_2}{x^0 Z^1 - x^1} \;.
\]
Combining these three expressions one gets the quadratic equation for $Z^1$:
\[
	(Z^1)^2 + \frac{y.x - 2y_1 x^1}{y_1 x^0 + x^2 x^3} Z^1 + \frac{y_2 y_3 - y_0 x^1}{y_1 x^0 + x^1 x^3} = 0 \;,
\]
where $y.x = y_I x^I$. Solving this we find an expression for $Z^1$ purely in terms of $x^I, y_I$:
\[
	Z^1 = -\frac{y.x - 2y_1 x^1}{2(y_1 x^0 + x^2 x^3)} \pm i\frac{\sqrt{W}}{2(y_1 x^0 + x^2 x^3)} \;,
\]
where 
\[
	W = -(y.x)^2 + 4y_1 x^1 y_2 x^2 + 4y_1 x^1 y_3 x^3 + 4 y_2 x^2 y_3 x^3 + 4 x^0 y_1 y_2 y_3 - 4 y_0 x^1 x^2 x^3 \;.
\]
By identical calculations, or simply by noting the symmetry between $Z^1, Z^2, Z^3$, we obtain similar expressions for $Z^2,Z^3$:
\begin{align*}
	Z^2 = -\frac{y.x - 2y_2 x^2}{2(y_2 x^0 + x^1 x^3)} \pm i\frac{\sqrt{W}}{2(y_2 x^0 + x^1 x^3)} \;, \\
	Z^3 = -\frac{y.x - 2y_3 x^3}{2(y_3 x^0 + x^1 x^2)} \pm i\frac{\sqrt{W}}{2(y_3 x^0 + x^1 x^2)} \;.
\end{align*}
Next, again by direct calculation one obtains the expression
\[
	\bar{Y}^0 = -\frac{2(x^0 Z^1 - x^1)}{Z^1 - \bar{Z}^1} \;,
\]
and, hence,
\[
	Y^0 \bar{Y}^0 = \frac{1}{W} \left( {x^0}^2 W + \left(x^0(y.x) + 2 x^1 x^2 x^3\right)^2 \right) \;.
\]
Also by direct calculation one can show that
\[
	(y_1 x^0 + x^2 x^3) (y_2 x^0 + x^1 x^3) (y_3 x^0 + x^1 x^2) = \tfrac{1}{4} \left( (x^0)^2 W + \left(x^0(y.x) + 2 x^1 x^2 x^3\right)^2 \right) \;.
\]
Substituting the above expressions into (\ref{eq:KahlerPotential}), we obtain
\[
	e^{\cal -K} = \pm 4 W^{1/2} \;.
\]
We now restrict ourselves to physically relevant configurations, where the RHS is strictly positive.
Since  $H = -\tfrac{1}{2}e^{{\cal -K}}$ we can write the Hesse potential explicitly in terms of $x^I,y_I$ as
\begin{align}	
	H(x,y) &= -2 \Big( -(y.x)^2 + 4y_1 x^1 y_2 x^2 + 4y_1 x^1 y_3 x^3 + 4 y_2 x^2 y_3 x^3 \notag \\
	&\hspace{14em} + 4 x^0 y_1 y_2 y_3 - 4 y_0 x^1 x^2 x^3 \Big)^{1/2} \;. 
	\label{eq:STU_Hesse_real}
\end{align}

One can use a similar procedure to determine the Hesse potential in terms of the imaginary parts of $Y^I,F_I$, which we denote by $u^I = \text{Im} (Y^I)$ and $v_I = \text{Im}(F_I)$.
What one obtains is precisely the same expression:
\begin{align}	
	H(u,v) &= -2 \Big( -(v.u)^2 + 4v_1 u^1 v_2 u^2 + 4v_1 u^1 v_3 u^3 + 4 v_2 u^2 v_3 u^3 \notag \\
	&\hspace{14em} + 4 u^0 v_1 v_2 v_3 - 4 v_0 u^1 u^2 u^3 \Big)^{1/2} \;. 
	\label{eq:STU_Hesse_imaginary}
\end{align}
The reason why we obtain the same result is that the Hesse potential is independent of the phase of $Y^I$, i.e. it is invariant under $U(1)$ transformations $Y^I \to e^{i\alpha} Y^I$. The imaginary parts of $Y^I, F_I$ are simply the real parts of $e^{-i\pi/2} Y^I, e^{-i\pi/2} F_I$, which describe the same Hesse potential.

\subsection{Hesse potential for models of form 
$F = \frac{f(Y^1,\ldots,Y^n)}{Y^0}$ \label{Hessepotential2}}

We now extend the previous discussion to models with a 
prepotential of the form
	\[
		F(Y) = \frac{f(Y^1,\ldots,Y^n)}{Y^0} \;.
	\]
	Since $F$ is a homogeneous function of degree 2 it follows that f is homogeneous function of degree 3.
In this case it is not possible to obtain an expression for the Hesse 
potential in closed form. However, one can still show that the Hessian
metric $\tilde{H}_{ab}$ takes the from (\ref{eq:Htilde2}) when restricting
to axion-free field configurations (\ref{eq:ax_free_condition}). The 
part of the Hessian metric relevant for this subspace can consistently
be obtained by setting half of the variables of the Hesse potential to
zero, and for this truncated Hesse potential we can obtain an 
explicit expression.

	Recall the definition of $x^I, y_I$ and $u^I, v_I$:
	\begin{align*}
		x^I + i u^I := Y^I &= \left( \begin{array}{c} Y^0 \\ Y^1 \\ \vdots \\ Y^n \end{array} \right) \;, \\
		y_I + iv_I := F_I &= \left( \begin{array}{c} -\frac{f(Y^1, \ldots, Y^n)}{Y^0{}^2} \vspace{.5em} \\ \frac{f_1(Y^1, \ldots, Y^n)}{Y^0} \\ \vdots \\ \frac{f_n(Y^1, \ldots, Y^n)}{Y^0} \end{array} \right) \;.
	\end{align*}
	
	We will now impose the conditions (\ref{eq:ax_free_condition}), which restrict us to the particular class of solutions for which $Y^A$ are purely 
imaginary, $Y^0$ is purely real and $F_0$ is purely imaginary. In this case the fields $x^I,y_I$ can be given explicitly in terms of $u^I,v_I$ by
	\begin{equation}
		\left( \begin{array}{c} x^0 \\ x^1 \\ \vdots \\ x^n \\ y_0 \\ y_1 \\ \vdots \\ y_n \end{array}\right)  = \left(\begin{array}{c} \lambda \\ 0 \\0 \\ 0 \\ 0 \\ -\frac{f_1(u^1, \ldots, u^n)}{\lambda} \\ \vdots\\ -\frac{f_n(u^1, \ldots, u^n)}{\lambda}  \end{array} \right)\;, \label{eq:xy_uv}
	\end{equation}
	where
	\[
		\lambda = -\sqrt{\frac{f(u^1, \ldots, u^n)}{v_0}} \;.
	\]
	One must choose the negative sign in the expression for $\lambda$ in order to ensure that the Hesse potential is strictly negative. The K\"ahler potential can be written as
	\[
		e^{\cal -K} = -i\left(Y^I \bar{F}_I - F_I \bar{Y}^I \right) = 8 \sqrt{v_0 f(u^1,\ldots,u^n)} \;,
	\]
	and since $e^{\cal -K} = -2 H$ we have the following explicit expression for the Hesse potential in terms of $u^I,v_I$:
	\begin{equation}
		H(u,v) = - 4 \sqrt{v_0 f(u^1,\ldots,u^n)} \;. \label{eq:axion_free_Hesse_uv}
	\end{equation}

	We would now like to find an equivalent expression for the Hesse potential in terms of $x^I, y_I$. Here we cannot use the same trick of making a $U(1)$ rotation as in the STU model, since imposing the conditions (\ref{eq:ax_free_condition}) implicitly brakes the $U(1)$ invariance of the system. Geometrically the condition selects a lower dimensional hypersurface which no longer has this $U(1)$ isometry. 
	
	Finding an explicit expression for the Hesse potential in terms of $x^I, y_I$ would involve inverting the relations (\ref{eq:xy_uv}), which in general cannot be calculated in closed form. However, we will now show that the 
Hesse potential can be consistently restricted to the subspace of 
axion-free solutions, where it separates into two distinct factors: 
		\begin{equation}
			H(x,y) = \sqrt{x^0} \, h(y_1, \ldots ,y_n) \;, \label{eq:axion_free_Hesse_xy}
		\end{equation}
		where $h$ is some homogeneous function of degree 3/2. This property is crucial in demonstrating the existence of non-BPS solutions to such models.
				
		Firstly, on the subspace of axion-free solutions, half
of the variables $x^I,y_I$ are zero. We denote the restricted Hesse 
potential by
		\[
			H(x,y) = H(x^0, y_1, \ldots, y_n)\;.
		\]
		Next, observe that for axion-free field configurations
		\[
		 	x^0 v_0 = -\sqrt{v_0 f(u^1,\ldots,u^n)} \;, \qquad \text{and} \qquad
		 	H = - 4\sqrt{v_0 f(u^1,\ldots,u^n)} \;,  
		\]
		\[
		 	\Rightarrow \;\; H = 4x^0 v_0 \;.
		\]
		Taking partial derivatives with respect to $x^0$ we find
		\[	
			\frac{\partial H}{\partial x^0} = 4v_0 + 4x^0 \frac{\partial v_0}{\partial x^0} \;.
		\]
Note that it does not make a difference whether we impose the axion-free
condition before or after taking derivatives with respect to $x^0$, because
the axion-free condition does not involve this variable. 
		But we know from (\ref{Ha}) that
		\[
			\frac{\partial H}{\partial x^0} = 2v_0 \;,
		\]
		and, hence, 
		\[
			x^0 \frac{\partial v_0}{\partial x^0} = -\frac{1}{2} v_0 \;, \hspace{3em} \Rightarrow \;\; v_0 = \frac{1}{\sqrt{x^0}} \frac14h(y_1, y_2, y_3) \;,
		\]
		for some specific, but as yet undetermined, function $h$.
		The restriction of the Hesse potential to axion-free configurations is therefore given by (\ref{eq:axion_free_Hesse_xy}). This allows us
to determine components of $\tilde{H}_{ab}$ which we need to go from
(\ref{eq:Htilde1}) to (\ref{eq:Htilde2}).

\providecommand{\newblock}{}

\end{document}